\documentclass[letterpaper, 12pt]{article}

\usepackage{amsmath,amssymb,mathrsfs}
\usepackage{graphicx}
\usepackage{color}
\usepackage{ushort}
\usepackage{url}
\usepackage{lscape}
\usepackage{subcaption}
\usepackage{pdflscape}
\usepackage{bbm}
\usepackage{epstopdf}
\usepackage{amsthm}
\usepackage[onehalfspacing]{setspace}
\usepackage[longnamesfirst]{natbib}
\usepackage{enumerate,enumitem}
\usepackage{libertine}
\usepackage[T2A, T1]{fontenc}
\usepackage[utf8]{inputenc}
\usepackage[english]{babel}
\usepackage{soul}
\usepackage{tcolorbox}
\usepackage{xcolor,colortbl}
\usepackage{threeparttable,multirow}
\usepackage{xfrac}
\usepackage{threeparttable}
\usepackage{rotating}
\usepackage{tabularx,ragged2e,booktabs}
\usepackage{subfiles}
\usepackage{amsmath}
\usepackage{amssymb}
\usepackage[margin=1in]{geometry}
\usepackage{bm}
\usepackage[title]{appendix}
\usepackage{setspace}
\usepackage{accents}
\usepackage{comment}

\definecolor{red}{rgb}{0.9,0.1,0.4}

\DeclareMathOperator{\supp}{supp}
\DeclareMathOperator{\E}{\mathbb{E}}
\DeclareMathOperator{\argmax}{argmax}

\DeclareMathOperator{\Cov}{Cov}
\DeclareMathOperator{\Corr}{Corr}

\DeclareMathOperator{\Std}{Std}

\DeclareMathOperator{\cl}{cl}

\newtheorem{proposition}{Proposition}

\newtheorem{lemma}{Lemma}
\newtheorem{theorem}{Theorem}
\newtheorem{assumption}{Assumption}

\newtheorem{definition}{Definition}

\newtheorem*{theorem**}{Proposition \theoremnum}
\newenvironment{theorem*}[1][]{
  \edef\theoremnum{\if\relax\detokenize{#1}\relax\else~#1\fi}
  \begin{theorem**}
}{
  \end{theorem**}
}

\definecolor{Blue}{RGB}{0,32,216}
\usepackage[colorlinks=true,allcolors=Blue]{hyperref}
\usepackage{xcolor}
\usepackage{rotating}
\title{Robust Contracting with Career Concerns\thanks{The authors are grateful to Dirk Bergemann, Hector Chade, Carlo Cusumano, Songzi Du, Joel Flynn, Robert Gibbons, Nima Haghpanah, Marina Halac, Daniel Hauser, Navin Kartik, Elliot Lipnowski, Giuseppe Moscarini, Ferdinand Pieroth, Daniel Rappoport, Doron Ravid, Evan Sadler, Philipp Strack, Nicholas Wu, and Kai Hao Yang for their helpful discussions and comments. We also thank the participants of the 36th Stony Brook conference and the seminars at Yale.}}
\author{Tan Gan\thanks{Tan Gan: Department of Management, London School of Economics (email: \href{mailto:t.gan2@lse.ac.uk}{t.gan2@lse.ac.uk}; address: Flat 28 Pied Bull Court, WC1A 2JR, London, UK).} \and Hongcheng Li\thanks{
        Hongcheng Li:  Department of Economics, Yale University (e-mail: \href{mailto:hongcheng.li@yale.edu}{hongcheng.li@yale.edu}; address: 27 Hillhouse Ave, B04, New Haven, CT 06511, USA).}}
                  
\date{May 25, 2026}

\begin{document}
\onehalfspacing
\maketitle

\begin{abstract}
    \noindent We study optimal contracting when workers face career concerns. Labor markets infer ability from performance, but effort affects how informative performance is. This feedback can generate strategic uncertainty: bonuses inducing effort under optimistic beliefs about effort may fail under pessimistic beliefs. We characterize this force through a criterion tied to skill-effort complementarity and solve for the least-cost policy implementing effort in every equilibrium. Under strategic uncertainty, the employer uses dispersed bonuses. High bonuses rule out pessimistic beliefs, raising the reputational stakes and letting lower bonuses motivate effort. Pay dispersion among observationally identical workers grows with career concerns and skill-wage assortativeness.
\end{abstract}



\newpage

\section{Introduction}

Career concerns and explicit contracts are two central sources of incentives in moral-hazard problems, yet their interaction remains underexplored. Career concerns arise whenever labor markets use a worker's past performance to assess his ability and shape his future compensation. Anticipating this inference, workers exert effort when the incentives provided by their employers, together with the reputational returns to good performance, are sufficient to cover their effort costs. Career concerns can therefore allow employers to reduce direct monetary compensation. At the same time, they make contracting more subtle: markets must interpret worker performance while forming beliefs about the incentives workers faced and their resulting effort choices.

This interaction between market beliefs and worker behavior can create strategic uncertainty for employers. When effort affects how informative performance is about ability, the market's expectations about effort also affect the worker's career-concern incentives. Consequently, a contract that induces effort when the market expects effort (and therefore views performance as informative) may fail to do so when the market instead expects shirking and draws different inferences from the same performance. Multiple equilibria can then arise, giving employers reason to design compensation that robustly induces effort across different market expectations.

The career-concerns literature largely avoids this multiple-equilibria problem. \citeauthor{holm1999}'s \citeyearpar{holm1999} seminal work isolates career concerns in a setting without explicit incentive contracts and with additive monitoring, while \cite{gimu1992} introduce optimal contracts but retain the same additive structure. Under additive monitoring, the informativeness of performance about ability is independent of effort. \cite{dejt1999b} show by example that multiplicative monitoring can generate multiple equilibria, but their analysis, like \citeauthor{holm1999}'s, abstracts from explicit contracting. The issue becomes particularly relevant when worker effort is jointly motivated by contracts and career concerns. A cost-minimizing employer reduces explicit incentives to the point where the worker is just willing to exert effort, so the contract sits at the margin where a small shift in market expectations can make shirking optimal. Consequently, in such environments, multiple equilibria arise as an intrinsic feature of optimal contracting. This leaves open two research questions: what criterion identifies the environments in which strategic uncertainty arises, and how should employers design compensation to induce effort robustly when it does arise?

To answer these questions, we study a model of contracting with career concerns. An employer hires a worker for a project whose outcome is either success or failure. The worker privately chooses whether to exert costly effort, and the probability of success depends on both his effort and his unobserved ability. As in standard career-concerns models, all players are initially uninformed about the worker's ability, and the labor market uses the project outcome to update its beliefs. The employer offers success-contingent bonuses and can commit to an \emph{incentive policy}, defined as a distribution over such bonuses. Equivalently, the firm commits to a compensation structure for a group of observationally identical workers, while each worker privately observes the particular contract offered to him. The labor market observes the firm's incentive policy and the worker's project outcome, but not the worker's realized contract.\footnote{This information asymmetry is consistent with many labor-market settings. Section \ref{contractual transparency} discusses the legal and regulatory background that motivates the assumption, including the active policy debate surrounding contractual privacy, and studies how the analysis changes as the labor market observes more information about individual contract realizations.} The worker's continuation payoff is then determined by the market's posterior belief about his ability. The employer's problem is to choose the least-cost incentive policy that induces effort in \emph{every} perfect Bayesian equilibrium.

We begin our analysis by examining two benchmarks. In the first, the employer evaluates policies by her preferred equilibrium, so she ignores strategic uncertainty and asks only that effort be sustained in that equilibrium. This partial-implementation benchmark yields a degenerate policy: the employer offers a single bonus that just makes the worker willing to exert effort when the market expects him to work. We call this bonus the \emph{partial-implementation bonus}. Strategic uncertainty is present when relying on this bonus leaves the employer exposed to another equilibrium in which the worker shirks with positive probability. In the second benchmark, the employer requires effort in every equilibrium but restricts attention to deterministic contracts. The resulting \emph{deterministic robust bonus} is pinned down by the most adverse market beliefs, that is, the beliefs that minimize career concerns. This bonus is therefore high enough to overcome any adverse market expectations.

Proposition \ref{proposition-benchmarks} shows that strategic uncertainty arises if and only if the partial-implementation bonus is strictly below the deterministic robust bonus. The contracting component leads to such a sharp criterion: the partial-implementation benchmark pushes work incentives to the margin by offering just enough to elicit effort when the market correctly anticipates work. At this margin, any market belief that lowers career concerns tips the worker toward shirking, so multiplicity arises precisely when such a belief exists.

To illustrate this condition, suppose that the project's success probability absent effort is independent of ability, while effort raises the success probability by more for higher-ability workers. If the market expects shirking, the outcome is uninformative about ability and generates no career-concern incentives. If instead the market expects effort, career concerns are present because success becomes a signal of high ability relative to failure. Strategic uncertainty arises because the worker's incentives are strongest precisely when the market expects him to work. More generally, Proposition \ref{claim 1} shows that strategic uncertainty arises under skill-effort complementarity: effort makes performance more informative about skill. In a linear environment, Proposition \ref{claim 2} shows that this condition is also necessary: strategic uncertainty is present if and only if worker productivity and labor-market pay are positively correlated. Since empirical evidence indicates substantial assortative matching between skills and wages,\footnote{See, for example, \cite{abkm1999}; \cite{cahk2013}; \cite{sopg2019}.} this criterion suggests that strategic uncertainty is likely to be relevant in many contracting environments.

Our main result, Theorem \ref{main result}, characterizes the robustly optimal incentive policy. We show that the employer uses randomization over bonuses if and only if strategic uncertainty is present. Specifically, under strategic uncertainty, the optimal policy randomizes over bonuses ranging from the partial-implementation bonus to the deterministic robust bonus. The highest bonus is large enough to induce effort regardless of career-concern incentives, thereby ruling out the most pessimistic market beliefs. Once such beliefs are ruled out, the market attaches greater reputational significance to performance, strengthening career concerns and allowing lower bonuses also to induce effort. Repeating this logic yields a policy that gradually lowers bonuses while assigning enough probability to higher bonuses to maintain strong career incentives. To minimize costs, the employer eliminates each potential low-effort belief only at the margin: given any such belief, the worker has just enough incentive to work. The characterization does not rely on global supermodularity. When the reputation effect is locally submodular, the optimal policy may include mass points.

The randomization has a natural interpretation. Consider a firm hiring a cohort of workers with similar observable qualifications. The firm may use a compensation system that allows offers to vary within a band, reflecting negotiation, recruiting timing, relocation needs, liquidity constraints, hiring-manager discretion, or other factors that are not fully observed by outside employers. Workers who receive high bonuses have strong direct incentives to exert effort. Because outside employers know that the firm's compensation system sometimes provides strong incentives, they expect greater effort in the cohort and interpret individual performance under a more demanding standard. Poor performance is then less easily attributed to weak incentives and more likely to be viewed as bad news about ability. This reputational pressure helps sustain effort even among workers who receive lower bonuses. In this way, dispersion in compensation can substitute for paying every worker the deterministic robust bonus.

Beyond its implications for optimal contracting, our results uncover a new mechanism for residual pay inequality: pay differences among workers with the same observable characteristics. Understanding such differences is a long-standing and central theme in labor economics.\footnote{\cite{mort2003} emphasizes that observationally similar workers often receive different contracts and that observable characteristics typically explain only a limited share of compensation variation.} In our framework, within-group dispersion arises as part of the employer's optimal policy to sustain effort under different market expectations. By varying compensation within a group of observationally identical workers, the employer shifts how outside markets interpret individual performance, and thereby the strength of workers' career-concern incentives.

To connect the mechanism to testable predictions, we derive comparative statics for when residual pay dispersion should be larger. We compare optimal policies using the \emph{de-meaned convex order}, which ranks pay distributions by dispersion after netting out differences in average pay and captures standard measures such as variance and pay range. Proposition \ref{proposition-occupational mobility} shows that dispersion increases when career concerns carry more weight in motivating effort. Thus, residual inequality should be greater in settings where future labor-market payoffs are more important, such as labor markets with greater occupational mobility. Proposition \ref{proposition-assortativeness skill premium} further shows that, in a linear environment, dispersion increases when worker productivity and labor-market pay are more strongly linked, either because more productive workers are more often matched to higher-paying jobs, or because the market places a larger premium on the skills that make effort more productive. These latter forces are typically associated with pay differences across skill groups; our results show that they can also generate residual pay dispersion among workers with the same observable characteristics. Intuitively, they make effort and career concerns more complementary, intensifying strategic uncertainty; employers respond optimally with greater dispersion in bonuses. The assortative-matching prediction is consistent with \citeauthor{cahk2013}'s \citeyearpar{cahk2013} evidence that greater worker-firm assortativeness contributed to rising wage inequality in West Germany.

Finally, we study three extensions of the model. The first contrasts our private-contracting environment with a public-contracting benchmark in which not only the worker but also the market observes the worker's realized contract. In that benchmark, the optimal incentive policy is degenerate at the deterministic robust bonus. Randomization is therefore useful only when individual contract terms are private and the employer wants to induce effort in every equilibrium. This finding speaks to an active policy debate on compensation privacy and inequality. Existing discussions emphasize how contractual privacy may reduce pay differences across observable groups. Our analysis highlights a different, within-group margin: by making individual contracts harder for outside employers to observe, contractual privacy can make optimal compensation more dispersed among observationally identical workers. We further analyze settings in which contract information partially leaks to the labor market.

The second extension considers a worker who is privately informed about his ability, in a labor market that values higher ability. Strategic uncertainty arises if and only if the gap between career incentives under working and shirking exceeds a threshold, again reflecting skill-effort complementarity. Proposition \ref{main result-private type} characterizes the robustly optimal incentive policy. The policy now combines a mass point at the partial-implementation bonus with continuous dispersion above it: the mass point sustains effort in the desired full-working equilibrium, while the dispersed higher bonuses induce higher-ability workers to work relatively more under adverse market expectations. This policy has a natural institutional interpretation: the employer publicly commits to a minimum compensation level and privately offers individualized bonuses to some workers.

The third extension studies a dynamic version of the model where the worker's labor-market continuation value arises endogenously from a future employer's contracting decision. We show that the first-period employer's problem reduces to that of the main model, with the career value now micro-founded by the future employer's contingent policy offers. Our main message therefore applies whether career concerns are specified through an exogenous continuation payoff or derived endogenously from subsequent contracting.

\emph{Outline.} Section \ref{model} presents the model and motivates the use of randomized incentive policies. Section \ref{strategic uncertainty} introduces the partial-implementation benchmark and the deterministic robust benchmark and uses them to establish a sharp criterion for strategic uncertainty. Section \ref{robustly optimal incentive policy} characterizes the robustly optimal incentive policy. Section \ref{interpretations and implications} interprets the criterion through special cases and derives comparative statics for residual pay dispersion. Section \ref{discussions} develops the three extensions. All proofs are in the Appendix.

\paragraph{Literature review.}
This paper contributes first to the literature on career concerns, pioneered by \cite{holm1999}. Relative to this literature, our contribution is to study explicit contract design when labor-market inference can generate equilibrium multiplicity, and to characterize the least-cost incentive policy that induces effort in every equilibrium. To our knowledge, ours is the first full-implementation analysis in the career-concerns literature. We show that dispersed bonuses can deliver explicit incentives that complement the implicit incentives generated by career concerns. Our analysis also accommodates flexible monitoring technologies, allowing effort to affect the informativeness of performance about ability in general ways.

\cite{dejt1999b} first recognized that multiplicative monitoring can generate multiple equilibria in career-concerns environments, but their analysis abstracts from explicit contracting. Other work studies richer contracting or learning environments while avoiding this multiplicity problem. \cite{gimu1992} study optimal contracts in a Holmstr\"{o}m-style Gaussian environment, where additive monitoring rules out multiplicity. \cite{boho2017} study dynamic career concerns with exponential learning and obtain uniqueness through the substitutability of effort across time. \cite{rodi2017} studies flexible monitoring in a non-contracting setting, focusing on the principal-preferred equilibrium. We instead combine explicit contracting with flexible monitoring and require effort to be induced across equilibria.

The paper is also related to the literature on contracting with externalities and adversarial equilibrium selection. One branch, following \cite{wint2004}, studies moral-hazard problems; another, following \cite{sega2003}, studies environments with bilaterally contractible agent decisions. Our paper belongs to the first branch. In this branch, \cite{hakw2024}, \cite{capo2025}, and \cite{cugp2023} study monitoring in teams, while \cite{moya2020}, \cite{halr2022}, and \cite{moot2024} study information provision as an incentive instrument. In the second branch, \cite{halr2024} study robust implementation with hidden types, which relates to our informed-worker extension; \cite{bewi2012} analyze coordinating multi-agent participation and discuss mixed externalities in an extension.\footnote{More broadly, the paper is related to the literature on full implementation in mechanism design, including \cite{OllarPenta17,OllarPenta23}, \cite{TE20251285}, and \cite{PeiStrulovici2024}.}

Most papers in this literature study externalities that run through the actions of multiple interacting agents. In our model, by contrast, the externality operates through the equilibrium beliefs of a third party: the outside labor market. In this respect, the closest papers are \cite{ahls2022} and \cite{gali2024}, which also feature externalities generated by a third party's beliefs. The differences are twofold. First, those papers study information-selling environments, where strategic externalities arise from unraveling under discretionary disclosure. We study contracting with career concerns, where the externality comes from workers' labor-market reputations. Second, their models involve bilaterally contractible decisions, whereas ours is a moral-hazard model with hidden effort. In addition, while much of the contracting-with-externalities literature focuses on complementarities, our framework allows the reputation effect to be locally complementary or substitutable.

Our contractual-transparency extension is related to \cite{halr2021}, but the object of privacy and the economic implications are different. They study intra-firm transparency: coworkers do not observe one another's contracts, and privacy leads to symmetric contracts among similar workers. We study inter-firm transparency: outside employers do not observe a worker's realized past contract. In our setting, contractual privacy can make optimal compensation more dispersed among observationally identical workers. Thus, the two papers differ both in the relevant audience for contract information and in the direction of the effect on pay dispersion.

Finally, the paper speaks to a broad literature on pay inequality. We identify a within-group mechanism for residual pay inequality: compensation dispersion among workers with the same observable characteristics can arise as part of a robustly optimal incentive policy. The comparative statics connect this dispersion to the importance of career concerns, occupational mobility, skill premia, and assortative matching between worker productivity and labor-market pay. In this respect, our analysis complements empirical work on pay dispersion and assortative matching, including \cite{abkm1999}, \cite{cahk2013}, and \cite{sopg2019}. It also complements recent theoretical work on labor-market beliefs and inequality. \cite{bags2024} study how monitoring determines whether early-career cross-group discrimination spirals or self-corrects. More broadly, \cite{boir2019} study how labor-market beliefs about workers' productivity evolve over time and generate persistent discrimination across demographic groups. Both papers study the dynamic evolution of cross-group disparities; we complement this line of inquiry with a within-group channel.

\section{Model}\label{model}

\paragraph{Setup.} A risk-neutral employer (she) hires a risk-neutral worker (he) to work on a project. The worker has ability $k \in K$, where $K$ is finite and contains at least two elements. Ability is drawn according to a full-support prior $\mu^0 \in \Delta(K)$. As in standard career-concerns models, we assume that $k$ is initially unobserved by both the employer and the worker, who share the same prior. The worker chooses whether to work at cost $c>0$ or shirk at no cost. If a worker of ability $k$ shirks, the project succeeds with probability $p_{0k} \geq 0$; if he works, the success probability rises to $p_{0k}+p_k\leq 1$, where $p_k>0$. We assume that both success and failure can occur on path: $p_{0k'}>0$ for some $k'$, and $p_{0k''}+p_{k''}<1$ for some $k''$. The project outcome is public and contractible, while the worker's effort choice is private and thus subject to moral hazard. It will be useful to define the \emph{effective cost}
\begin{equation}\label{effective cost}
\lambda:=\frac{c}{\E_0[p_k]}=\frac{c}{\sum_{k\in K}\mu^0_k p_k},
\end{equation}
where $\E_0$ denotes expectation under the prior belief. Thus, $\lambda$ is the effort cost normalized by the ex ante productivity gain from work.

\paragraph{Career concerns.} After the project is completed, the worker leaves the current employer and enters the labor market. The market values the worker's ability and initially holds the prior. We represent the worker's \emph{post-employment value} by a function $v:\Delta(K)\to \mathbb R_+$, where $v(\mu)$ is the transfer paid to the worker when the market's posterior belief about his ability is $\mu$. For our main characterization, we require only that $v(\cdot)$ be continuous in the Euclidean norm.

\paragraph{Incentive policy.} We assume that the worker is protected by limited liability, so it is without loss to focus on success-contingent bonuses; that is, the employer pays the worker only upon project success. We therefore identify a contract with a nonnegative bonus and use these two terms interchangeably. The employer commits to an incentive policy $F\in\Delta(\mathbb R_+)$, a distribution over such bonuses. Each $b\geq 0$ drawn from $F$ is a realized contract under which the employer pays bonus $b$ upon success.

For a concrete interpretation of $F$, imagine that the employer hires many ex-ante identical workers and assigns each a contract drawn from $F$, with individual offers hidden from the market. Choosing $F$ is thus equivalent to choosing a contract structure for these observably identical workers. Because the employer cannot price ability, the same policy applies to all workers; our setup thus suppresses cross-group contract dispersion, and any non-degenerate $F$ directly represents \emph{within-group} inequality.

Such dispersion admits several natural micro-foundations. First, the firm may institutionalize bilateral bargaining within a posted range to accommodate workers' idiosyncratic non-productivity factors such as relocation costs, family circumstances, or personal liquidity needs. Second, it may delegate offer-setting to hiring managers, whose decisions reflect productivity-irrelevant factors such as affinity, demeanor, or cultural fit that are unobserved by the market. Third, it may design institutions that let external conditions (labor-market tightness, budget cycles, or recruiting urgency) pass through to realized contracts. In each case, the employer chooses $F$ by designing the associated institutions.

Our model also assumes that the market cannot observe the realized contract. As discussed in Section \ref{contractual transparency}, this asymmetric information reflects an active legal and regulatory environment that restricts firms' access to workers' past contracts; that section also considers four cases in which contract information partially leaks to the market.

\paragraph{Timing and payoffs.} Holding the prior belief, the employer \emph{publicly} offers an incentive policy $F\in\Delta(\mathbb{R}_+)$. The worker's ability $k$ is drawn from $\mu^0$ but not observed by anyone. The worker \emph{privately} learns the bonus level, say $b$, which is a realization of the policy $F$. Given this information, the worker \emph{privately} chooses whether to work ($a=1$), shirk ($a=0$), or randomize. Based on worker effort and ability, the project outcome $x$ is \emph{publicly} and stochastically realized, being either success ($x=1$) or failure ($x=0$). At this point, the market observes only the incentive policy and the project outcome, but not ability, the realized contract, or the worker's effort. Subsequently, the market forms its posterior belief $\mu$ via Bayes' rule and pays the worker the post-employment value $v(\mu)$.

The worker's ex-post payoff includes the working cost, the realized contract, and the post-employment value: $-ac+xb+v(\mu)$. The employer's objective will be specified below.

\paragraph{Equilibrium.} Each incentive policy $F$ induces a dynamic Bayesian game played by the worker and the market. We consider the set of all perfect Bayesian equilibria of the game. Specifically, an equilibrium is a tuple $g=(\sigma,\overline{\mu},\underline{\mu})$, where $\sigma:\mathbb{R}_+\rightarrow[0,1]$ denotes the worker's probability of working upon receiving a bonus realization $b\in\mathbb{R}_+$, and $\overline{\mu}$ and $\underline{\mu}$ are the market's posterior beliefs following project success and failure, respectively.

A perfect Bayesian equilibrium satisfies: (i) given the market's beliefs, for every bonus realization $b$, the worker maximizes his expected payoff by deciding whether to work
\begin{equation}\label{informed optimality}
    \begin{aligned}
        &\sigma(b)\in\argmax_{\widetilde{\sigma}\in[0,1]}\widetilde{\sigma}(V^1-V^0),\text{ where} \\
        &V^1=-c+\E_0[(p_{0k}+p_k)(b+v(\overline{\mu}))+(1-p_{0k}-p_k)v(\underline{\mu})],\text{ and} \\
        &V^0=\E_0[p_{0k}(b+v(\overline{\mu}))+(1-p_{0k})v(\underline{\mu})];
    \end{aligned}
\end{equation}
And (ii) given the worker's total working probability $q:=\E_F[\sigma(b)]=\int_{\mathbb{R}_+}\sigma(b)dF(b)$ pinned down by his strategy, the market's beliefs satisfy Bayes' rule: $\overline{\mu}$ and $\underline{\mu}$ assign each type $k$ the following probabilities
\begin{equation}\label{bayes rule}
    \begin{aligned}
        &\overline{\mu}_k=\frac{\mu^0_k[q(p_{0k}+p_k)+(1-q)p_{0k}]}{\sum_{j\in K}\mu^0_j[q(p_{0j}+p_j)+(1-q)p_{0j}]},\text{ and} \\
        &\underline{\mu}_k=\frac{\mu^0_k-\mu^0_k[q(p_{0k}+p_k)+(1-q)p_{0k}]}{1-\sum_{j\in K}\mu^0_j[q(p_{0j}+p_j)+(1-q)p_{0j}]}.
    \end{aligned}
\end{equation}
Bayes' rule applies because, by assumption, success and failure are always on path. When no confusion arises, we refer to the tuple $g=(\sigma,\overline{\mu},\underline{\mu})$ defined above simply as an \emph{equilibrium}. In addition, let $\mathcal{E}(F)$ collect all the equilibria induced by $F$. If $F$ is degenerate at some bonus level $b$, we also denote the equilibrium set simply by $\mathcal{E}(b)$. \\

\paragraph{Robustness.} We investigate the robustly optimal incentive policy that minimizes the employer's expected cost while inducing the worker to work with probability one in all equilibria.
Focusing on full working is without loss when the employer values success sufficiently more than failure. Inducing any interior working probability is analogous, as we show in Section \ref{main result-section}.

More formally, an incentive policy $F$ \emph{fully implements full working} if, in every equilibrium $g\in\mathcal{E}(F)$, the worker works with probability $q=\E_F[\sigma(b)]=1$. We collect all such incentive policies in $\mathcal{F}^{FI}$. This set is nonempty because the employer can always guarantee full working by offering a sufficiently high bonus. We can thus define the employer's objective

\begin{definition}\label{definition-solution concept}
    \emph{The employer's }minimal cost guarantee\emph{ for fully implementing full working is
    \begin{equation}\label{minimal cost guarantee}
        W^*=\inf_{F\in\mathcal{F}^{FI}}\sum_{k\in K}\mu^0_k(p_{0k}+p_k)\E_F[b].
    \end{equation}
    An incentive policy $F^*$ is }robustly optimal\emph{ if there is a sequence of incentive policies $(F_n)_{n=1}^{\infty}$ such that:
    \begin{enumerate}[nolistsep]
        \item The sequence $(F_n)_{n=1}^{\infty}$ is contained in $\mathcal{F}^{FI}$ and weakly converges to $F^*$;
        \item The cost guarantee, $\sum_{k\in K}\mu^0_k(p_{0k}+p_k)\E_{F_n}[b]$, converges to $W^*$.
    \end{enumerate}}
\end{definition}

Note that in (\ref{minimal cost guarantee}), the term $\sum_{k\in K}\mu^0_k(p_{0k}+p_k)$ is the total probability that a bonus is paid when the worker works with probability one. This constant is independent of the incentive policy, so the employer's essential objective is to minimize the expected bonus $\E_F[b]$.

\section{Strategic Uncertainty}\label{strategic uncertainty}

When does strategic uncertainty become a problem if the employer designs the contract while ignoring the possibility of multiple equilibria? We answer this question by examining two benchmarks, the partial-implementation benchmark and the deterministic robust benchmark, each relaxing one model element.

In the partial-implementation benchmark, the employer picks an incentive policy to partially implement work, that is, she ignores strategic uncertainty and chooses her preferred equilibrium. The cost-minimizing design here is called the \emph{partial-implementation policy}. In the deterministic robust benchmark, the employer fully implements full working but can use only a deterministic contract rather than a distribution. Her optimal choice is called the \emph{deterministic robust bonus}. We define the two benchmarks formally
\begin{equation*}
    \begin{aligned}
        \text{[Partial implementation]: }&\min_{F\in \Delta(\mathbb{R}_+)}\E_F[b]\text{, s.t. }\exists g=(\sigma,\overline{\mu},\underline{\mu})\in\mathcal{E}(F), \E_F[\sigma(b)]=1; \\
        \text{[Deterministic robust]: }&\inf_{b\geq0}b\text{, s.t. }\forall g=(\sigma,\overline{\mu},\underline{\mu})\in\mathcal{E}(b), \sigma(b)=1.
    \end{aligned}
\end{equation*}

\paragraph{Career value.} The analysis starts by defining the \emph{career value} $D:[0,1]\rightarrow\mathbb{R}$ as the difference between the post-employment value for project success and that for project failure. In (\ref{bayes rule}), the worker's total working probability $q$ uniquely pins down the belief system through Bayes' rule, so we rewrite the beliefs as two functions: $\overline{\mu}(q)$ and $\underline{\mu}(q)$. The career value is defined by
\begin{equation}\label{career value}
    D(q):=v(\overline{\mu}(q))-v(\underline{\mu}(q)).
\end{equation}
This concept yields a compact restatement of our equilibrium condition: by rewriting (\ref{informed optimality}), we find that the worker's strategy $\sigma$ forms an equilibrium if, given $q=\E_F[\sigma(b)]$, for all bonuses $b\geq0$
\begin{equation}\label{equilibrium concept}
    \sigma(b)\in\argmax_{\widetilde{\sigma}\in[0,1]}\widetilde{\sigma}[b+D(q)-\lambda],
\end{equation}
where the effective cost $\lambda$ is defined in (\ref{effective cost}). In other words, the worker decides by comparing his effective cost $\lambda$ and his effective gain $b+D(q)$, which includes both the explicit incentives $b$ provided in the contract and the implicit incentives of career concerns $D(q)$. Moreover, the career value $D(q)$ is produced by the market's correct expectation about the worker's total effort $q$.

Because $v(\cdot)$ is continuous and the beliefs $\overline{\mu}_k$ and $\underline{\mu}_k$ in (\ref{bayes rule}) are continuous in $q$, the career value $D(\cdot)$ is continuous in $q$. Note that even when $v(\cdot)$ exhibits an intuitive pattern, such as increasing in the probability of higher ability levels, $D(\cdot)$ is in general nonmonotonic, so the market's optimism about effort can both enhance and reduce work incentives. This requires us to handle an externality structure that can arbitrarily mix supermodularity and submodularity.

Finally, throughout the paper we make the non-essential assumption that career concerns never fully compensate for the working cost, that is, $\max_{q}D(q)<\lambda$. This ensures that the employer necessarily offers positive bonuses as explicit incentives in her optimal design.

\paragraph{Benchmark contracts.} To derive solutions, we first highlight the following two critical bonus levels
\begin{equation*}
    \underline{b}=\lambda-D(1);\quad\quad\quad\overline{b}=\lambda-\min_{q\in[0,1]}D(q).
\end{equation*}

Recall that when the market expects the worker to work with total probability $q\in[0,1]$, the worker faces career value $D(q)$. The equilibrium condition (\ref{equilibrium concept}) then implies that the bonus $b_q:=\lambda-D(q)$ makes the worker indifferent between working and shirking given the market expectation $q$. Moreover, if the incentive policy offers bonuses no less than $b_q$ with exactly the probability $q$, then such a market expectation is self-fulfilling in equilibrium.

The result below establishes that dispersion is not useful for partial implementation, in which case the employer faces the multiplicity issue precisely when the two benchmarks' solutions diverge.

\begin{proposition}\label{proposition-benchmarks}
    The deterministic robust bonus is $\overline{b}$, whereas the partial-implementation policy is degenerate at $\underline{b}$. Moreover, the partial-implementation policy is not robustly optimal if and only if $\underline{b}<\overline{b}$, or equivalently,
    \begin{equation}\label{criterion for strategic uncertainty}
        D(1)>\min_{q\in[0,1]}D(q).
    \end{equation}
\end{proposition}

Proposition \ref{proposition-benchmarks} delivers two messages. First, despite her ability to offer a bonus distribution, the partially implementing employer still uses a single bonus, creating no contract dispersion.
Second, if the employer ignores the underlying issue of multiplicity and thus offers $\underline{b}$ (or any policies approximating it), her goal is jeopardized by an undesirable equilibrium whenever criterion (\ref{criterion for strategic uncertainty}) holds. 
This criterion roughly states that if a higher effort $q$ tends to increase the gap $D$ between the worker's market pay after success and after failure, then the employer faces strategic uncertainty and needs to adopt a robust objective when designing the contract. Accordingly, we say that \emph{strategic uncertainty is present} (under partial implementation) if criterion (\ref{criterion for strategic uncertainty}) holds. We return to criterion (\ref{criterion for strategic uncertainty}) in Section \ref{interpretations and implications}, providing more interpretable sufficient conditions on model primitives.

To understand why offering $\underline b$ can fail, we explain how multiple equilibria emerge in our setting. The key lies in the interdependence between the worker's work decisions and the market's beliefs. On the one hand, as long as the offered bonuses are no less than $\underline{b}$ (equal to $b_q$ with $q=1$), the full-working equilibrium exists. Thus, the partial-implementation policy assigns all probability to this lower bound $\underline{b}$. On the other hand, this bonus does not guarantee full implementation because other market expectations may also be self-fulfilling. For example, if there is some $q<1$ with $D(q)=D(1)$, we have $\underline{b}=b_q$, so the career value $D(q)$ makes it optimal for the worker to randomize at the partial-implementation bonus (with working probability $q$). An undesirable mixed-strategy equilibrium thereby emerges. The second possibility arises when $D(1)>D(0)$, in which case $\underline{b}$ is lower than $b_q$ with $q=0$, so the career value $D(0)$ drives the worker to shirk at the partial-implementation bonus. Therefore, full shirking forms an undesirable equilibrium in this case. These undesirable equilibria vanish only when $D(1)<D(q)$ for all $q<1$. Furthermore, to fully implement $q=1$, the deterministic robust bonus must avoid the previous two cases and thus exceed $b_q$ for all $q<1$, so it is the infimum of such bonuses, $\overline b$.

Whenever criterion (\ref{criterion for strategic uncertainty}) holds, the deterministic robust bonus $\overline{b}$ strictly exceeds the partial-implementation bonus $\underline{b}$, leaving room for randomization to reduce the expected cost. The next section develops the analytical framework and characterizes the robustly optimal incentive policy.

\section{Optimal Incentive Policy}\label{robustly optimal incentive policy}

Section \ref{an auxiliary problem} reduces the main problem (\ref{minimal cost guarantee}) to an auxiliary problem that isolates the key constraints introduced by our robust focus. Section \ref{main result-section} presents and interprets the main result, Theorem \ref{main result}, including the intuition for how iteratively introducing contract dispersion achieves full implementation.

\subsection{An Auxiliary Problem}\label{an auxiliary problem}

This section establishes the equivalence between the employer's problem (\ref{minimal cost guarantee}) and an auxiliary problem. The auxiliary problem identifies two constraints, one for each of the employer's two main goals. First, the employer must ensure the desired full-working equilibrium exists. Second, the incentive policy must destabilize all undesirable equilibria for full implementation.

The first constraint follows directly from our arguments in Section \ref{strategic uncertainty} that, to keep the full-working equilibrium, the offered bonuses must be no less than the partial-implementation bonus, $\underline{b}$. Formally, any incentive policy $F$ must satisfy the following \emph{equilibrium-keeping} constraint
\begin{equation}\label{equilibrium keeping}
    \supp(F)\geq\underline{b}. \tag{EK}
\end{equation}

The second constraint introduces a series of conditions for breaking all bad equilibria. In every bad equilibrium, the worker's total working probability $q$ is strictly lower than one. According to the equilibrium condition (\ref{equilibrium concept}), the worker in equilibrium must play a threshold strategy. That is, there exists some bonus level $b$ such that he works [shirks] if the bonus realization is higher [lower] than $b$, and he is indifferent at $b$. Hence, every such equilibrium specifies a $q$ and a $b$, and these parameters are consistent with the same worker strategy if and only if
\begin{equation*}
    q\in[1-F(b^+),1-F(b^-)]\text{, which we also denote }b\in F^{-1}(1-q).
\end{equation*}

On the one hand, when the market expects the worker to work with a total probability $q$, the career value is $D(q)$. On the other hand, to rationalize the threshold bonus $b$, (\ref{equilibrium concept}) requires that the career value equal $\lambda-b$. Thus, this equilibrium does not exist if and only if the two career values differ
\begin{equation*}
    D(q)\neq\lambda-b.
\end{equation*}
In particular, if $D(q)>\lambda-b$, the market's expectation is too ``aggressive'', creating such a high career value that the worker wants to deviate to work more than expected. Likewise, when $D(q)<\lambda-b$, the ``conservative'' market expectation induces the worker to deviate toward more shirking.

However, despite these two possible ways of breaking equilibrium, we will show that an incentive policy that fully implements full working must break those bad equilibria by inducing ``aggressive'' market expectation alone. Formally, it must satisfy the following \emph{equilibrium-breaking} constraint
\begin{equation}\label{equilibrium breaking}
    \text{For all }q\in[0,1)\text{ and }b\in F^{-1}(1-q)\text{, }b+D(q)\geq\lambda. \tag{EB}
\end{equation}
To understand why equilibrium elimination must be directional, we first examine a candidate equilibrium with a very high threshold bonus $b_H$ and thus zero working probability $q_H=0$, for which (\ref{equilibrium breaking}) holds because $D(0)$ is bounded. Therefore, had there been some $q_L$ and $b_L$ with $b_L\in F^{-1}(1-q_L)$ such that $b_L+D(q_L)<\lambda$, the continuity of $D(\cdot)$ would let us construct $b'\in[b_L,b_H)$ and some associated $q'<1$ that form a bad equilibrium. This, however, violates full implementation. That is,
\begin{equation*}
\left.
\begin{aligned}
b_H+D(q_H) &> \lambda \\
b_L+D(q_L) &< \lambda
\end{aligned}
\right\}\Rightarrow\text{there exists }b'+D(q')=\lambda.
\end{equation*}

Crucially, the two constraints are also sufficient for finding the robustly optimal incentive policy.

\begin{proposition}\label{proposition-auxiliary problem}
    $F^*$ is robustly optimal if and only if it solves the following problem
    \begin{equation}\label{auxiliary problem}
        \min_{F\in\Delta(\mathbb{R_+})}\int_{\underline{b}}^{\infty}(1-F(b))db\qquad\text{s.t. (\ref{equilibrium keeping}) and (\ref{equilibrium breaking})}.
    \end{equation}
\end{proposition}

Recall that our setting incorporates arbitrarily mixed externalities because career value may be nonmonotonic in the market's expectation. Proposition \ref{proposition-auxiliary problem} translates the employer's robustness problem into a standard cost-minimization problem, with the (\ref{equilibrium breaking}) constraint restricting the direction in which the employer breaks all bad equilibria. This makes the general characterization possible. We later derive analogous results for related problems with partial working and an informed worker.

\subsection{Main Result}\label{main result-section}

Solving the auxiliary problem in Proposition \ref{proposition-auxiliary problem} yields our main characterization.

\begin{theorem}\label{main result}
    To induce full working, the unique robustly optimal incentive policy $F^*$ satisfies
    \begin{equation*}
        \begin{aligned}
            \supp(F^*)&=[\underline{b},\overline{b}]\text{, and for all }b\in[\underline{b},\overline{b})\text{, we have} \\
            F^*(b)&=1-\overline{q}(b^+)\text{, where }\overline{q}(b):=\max\{q:b+D(q)=\lambda\}.
        \end{aligned}
    \end{equation*}
    Moreover, $F^*$ is continuous if and only if $D(\cdot)$ is strictly increasing.\footnote{We take the right limit $\overline{q}(b^+)$ simply to maintain a right-continuous $F^*(\cdot)$.}$^,$\footnote{In addition to the full-working case, if the employer fully implements a fixed total working probability $Q\in[0,1)$, the result is characterized by $\supp(F^*)=\{0\}\cup[\underline{b}_Q,\overline{b}_Q]$ (with $Q<1$, $F^*$ has probability $1-Q$ to fire the worker by offering 0) and $F^*(b)=1-\overline{q}(b^+)$ for all $b\in[\underline{b}_Q,\overline{b}_Q]$, where $\underline{b}_Q:=\lambda-D(Q)$ and $\overline{b}_Q:=\lambda-\min_{q\in[0,Q]}D(q)$.}
\end{theorem}

Theorem \ref{main result} thus delivers a sharp prediction: criterion (\ref{criterion for strategic uncertainty}), namely $\underline{b}<\overline{b}$, governs \emph{both} whether strategic uncertainty is present and whether the robustly optimal policy exhibits contract dispersion.

To build intuition for how contract dispersion reduces costs in practice, think of an employer hiring a cohort of workers with similar backgrounds and observable qualifications. Rather than offering everyone the same high bonus, the employer deliberately creates contract dispersion, for example, by assigning different job titles or bonus packages that are only imperfectly linked to underlying ability. Workers who receive especially favorable treatment have strong direct monetary incentives to exert high effort. Their performance, in turn, shapes the firm's reputation in the external labor market: future employers come to expect that capable workers from this firm typically perform well. This expectation raises the reputational stakes for all employees. Consequently, even workers who receive less favorable treatment are compelled to work hard, because poor performance is more likely to be interpreted as evidence of low ability rather than weak incentives. By securing a small group of highly motivated, well-paid workers who anchor optimistic market beliefs, the employer can elicit effort from the remaining workers at lower bonuses. In this way, artificially created contract dispersion helps to rule out low-effort beliefs and substitutes for uniformly high pay.

To make the iterative construction concrete, we first record below a lemma that provides the career-value guarantee used in defining $F^*$.

\begin{lemma}\label{lemma-strict increasing}
    For all $\widetilde{b}\in(\underline{b},\overline{b})$ and $\widetilde{q}:=\max\{q':\widetilde{b}=\lambda-D(q')\}$, we have $D(q)>D(\widetilde{q})$ for all $q>\widetilde{q}$.
\end{lemma}

Lemma \ref{lemma-strict increasing} says that guaranteeing the market's expectation of effort at $\widetilde{q}$ bounds the worker's career concerns below by $D(\widetilde{q})$, so the bonus $\widetilde{b}$ suffices to incentivize effort. To illustrate the iterative construction in its simplest form, consider, for a small $\epsilon>0$, the following binary-bonus incentive policy $\widetilde{F}$, which improves upon the deterministic robust bonus $\overline{b}$
\begin{equation*}
    \widetilde{F}\text{ offers }\overline{b}+\epsilon\text{ with probability }\widetilde{q}\text{ and offers }\widetilde{b}+\epsilon\text{ with probability }1-\widetilde{q}.
\end{equation*}
Because $\widetilde{b}<\overline{b}$, for small $\epsilon$ the distribution $\widetilde{F}$ has a lower expected cost than the deterministic robust bonus. To see why $\widetilde{F}$ also fully implements full working, note first that the desired full-working equilibrium exists because both $\widetilde{b}$ and $\overline{b}$ are no lower than $\underline{b}$. The remaining undesirable equilibria fall into two categories: (i) the worker works with some probability $q<1$ at the high bonus while fully shirking at the low bonus; (ii) the worker fully works at the high bonus and works with some probability $q<1$ at the low bonus. The high bonus $\overline{b}+\epsilon$ is above the deterministic robust bonus, so working is dominant there regardless of career concerns: case (i) is eliminated. Anticipating this, the market expects the worker to work with probability at least $\widetilde{q}$, which by Lemma \ref{lemma-strict increasing} brings the career value to at least $D(\widetilde{q})$. Consequently, the low bonus $\widetilde{b}+\epsilon$ exceeds $\widetilde{b}=\lambda-D(\widetilde{q})$, which is high enough to make working optimal given the career-value guarantee $D(\widetilde{q})$: case (ii) is also ruled out. Of course, this two-point example is the simplest instance of the iterative construction that defines $F^*$ in Theorem \ref{main result}. Repeating the construction with finer bonus splits leads, in the limit, to the full support $[\underline{b},\overline{b}]$ of $F^*$.

What does $\overline{q}(\cdot)$ in Theorem \ref{main result} capture? Lemma \ref{lemma-strict increasing} provides the answer. If the employer commits a mass of at least $\overline{q}(b)$ to bonuses above $b$, the lemma guarantees a career value bounded below by $D(\overline{q}(b))$, and the defining identity $b+D(\overline{q}(b))=\lambda$ then makes the worker just willing to work at $b$. Hence $\overline{q}(b)$ is the smallest mass above $b$ that the employer must allocate to make the worker's incentive constraint bind at $b$. Setting $F^*(b)=1-\overline{q}(b)$ thus makes the (\ref{equilibrium breaking}) constraint bind pointwise
\begin{equation*}
    \text{For all }b\in[\underline{b},\overline{b}),\quad b+D(1-F^*(b))=\lambda.
\end{equation*}
This construction (approximately) eliminates each low-effort equilibrium at the margin. When the market expects a threshold strategy at $b$, the beliefs induced by $F^*$ generate a career value $D(1-F^*(b))=\lambda-b$. An equilibrium with this expectation is ruled out under perturbation: any small mass shift from below $b$ to above $b$ further raises the career value above $\lambda-b$ (by Lemma \ref{lemma-strict increasing}), luring the worker to work at bonuses just below $b$. We thereby construct a sequence of fully implementing policies that approximates $F^*$, as demanded by Definition \ref{definition-solution concept}. Intuitively, the employer assigns the smallest probability to bonuses above each threshold expected by the market, while ensuring that the expectation is only slightly more ``aggressive'' than the self-fulfilling level.

\begin{figure}[h]
    \centering
    \includegraphics[width=0.8\linewidth]{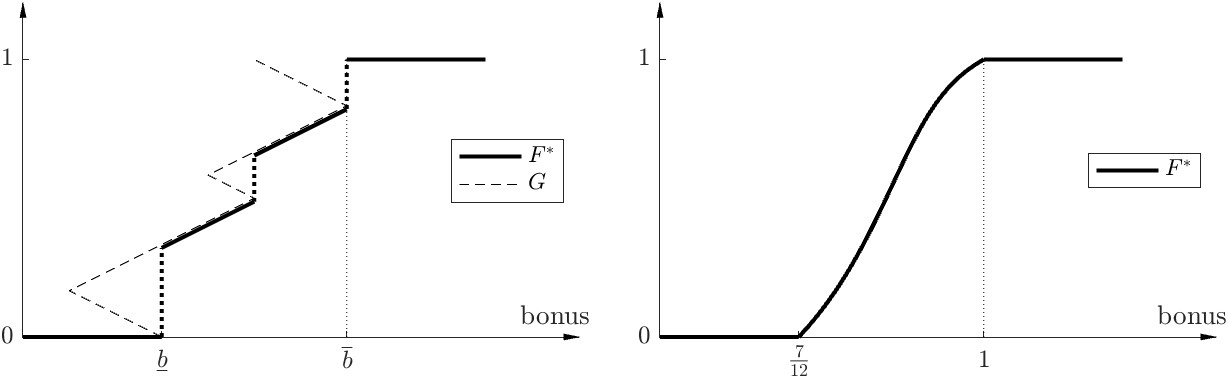}
    \caption{Robustly optimal incentive policy $F^*$. In the left panel, $G=\{(b,F):F=1-q\text{ and }b=\lambda-D(q)\text{ for some }q\in[0,1]\}$. The right-panel example considers linear market valuation, parameterized by $ K=\{H,L\}$ with prior $\mu^0_H=\mu^0_L=0.5$, post-employment value $v(\mu)=\mu_H$, shirking success probabilities $p_0=p_{0H}=p_{0L}=0.1$, additional success probabilities due to working $p_H=0.7$ and $p_L=0.3$, and effective cost $\lambda=1$. This gives $D(q)=\frac{10q}{(1+5q)(9-5q)}$, $\overline{b}=1-D(0)=1$, and $\underline{b}=1-D(1)=\frac{7}{12}$.}
    \label{fig-robustly optimal incentive policy}
\end{figure}

Figure \ref{fig-robustly optimal incentive policy} depicts the robustly optimal incentive policy in two examples. The left panel demonstrates the geometric construction of $F^*$. The graph $G$ consists of all pairs $(b,F)$ with $F=1-q$ and $b=\lambda-D(q)$ for some $q\in[0,1]$. $G$ generally exhibits a zigzag pattern because the induced career value function can be nonmonotonic. Theorem \ref{main result} states that the robustly optimal incentive policy $F^*$ is given by the highest nondecreasing curve lying below $G$. Furthermore, this example illustrates that mass points arise precisely when the career value is not strictly increasing, or in other words, when skill and effort become ``local substitutes''. The right panel examines a special linear-value case (the parameters are specified in the caption, and the linear environment is studied more systematically in Section \ref{interpreting the criterion}). In this case, the career value $D(q)$ is strictly increasing, so $F^*$ is continuous and free of mass points.

\section{Interpretations and Implications}\label{interpretations and implications}

This section interprets our results and draws out their implications. Section \ref{interpreting the criterion} translates criterion (\ref{criterion for strategic uncertainty}) into model primitives through three special cases: skill-effort complementarity, the linear environment, and the classic monitoring system. Section \ref{implications} then develops comparative statics that connect $F^*$ to several stylized empirical patterns of residual pay inequality.

\subsection{Interpreting the Criterion}\label{interpreting the criterion}

Although criterion (\ref{criterion for strategic uncertainty}) is necessary and sufficient for both strategic uncertainty and contract dispersion, it is stated in reduced form. To make it interpretable, we discuss three special cases of our model. Now that Theorem \ref{main result} is in hand, each interpretation simultaneously describes when strategic uncertainty arises and when the optimal policy exhibits dispersion.

\paragraph{Skill-Effort Complementarity.} What market features drive strategic uncertainty? Intuitively, criterion (\ref{criterion for strategic uncertainty}) holds when skill and effort are complements in the reputation effect of career concerns: greater effort makes performance more informative about skill. Specifically, if the workers whose effort raises the success probability more also receive better pay in the labor market, then increasing effort widens the pay differential between success and failure, providing stronger career-concerns incentives.

More concretely, we first impose a weak assumption on the post-employment value $v(\cdot)$ that the market pays the worker. We label the ability levels as integers $ K=\{1,2,...,K\}$ with $K\geq 2$,\footnote{Many examples we give in this paper, like the current one, abuse the notation $K$ and let it also represent its cardinality. We always assume $K\geq 2$ in these settings.} and suppose that the ability levels are ordered by the market's ``willingness-to-pay'' for each of them. Specifically, let $v(\cdot)$ be \emph{strictly increasing in first-order stochastic dominance}: for all $\mu^1,\mu^2\in\Delta( K)$,
\[\sum_{j=1}^k\mu^1_j\leq\sum_{j=1}^k\mu^2_j\text{ for all $k$, and some inequality is strict}\Rightarrow v(\mu^1)>v(\mu^2).\]
In other words, the market pays more if it believes the worker is more likely to have higher ability.

To fix ideas, consider a simple example: the success probability under shirking, $p_{0k}$, does not change with $k$, whereas the incremental success probability from working, $p_k$, is strictly increasing in $k$. This case assumes co-movement between skill value $v(\cdot)$ and effort effectiveness $p_k$. While no inference can be made from work outcome when the worker shirks, monitoring becomes informative only when the worker works. Consequently, criterion (\ref{criterion for strategic uncertainty}) holds here and strategic uncertainty is present.

We now generalize the example above to identify a sufficient condition for (\ref{criterion for strategic uncertainty}). To do so, for each ability level $k$, define the worker's ``effort effect'' on project outcome as $P_k$, and his ``shirking effect'' on project success and project failure as $Q^S_k$ and $Q^F_k$, respectively, as follows
\[P_k:=\frac{\mu_k^0p_k}{\sum_{j=1}^K\mu_j^0p_j};\quad\quad\quad Q_k^S:=\frac{\mu_k^0p_{0k}}{\sum_{j=1}^K\mu_j^0p_{0j}};\quad\quad\quad Q_k^F:=\frac{\mu_k^0(1-p_{0k})}{\sum_{j=1}^K\mu_j^0(1-p_{0j})}.\]
Let $P$, $Q^S$, and $Q^F$ be the vectors of these effects across ability levels, so that $P,Q^S,Q^F\in\Delta(K)$. Note that $Q^S$ and $Q^F$ are the posterior beliefs of a market that anticipates shirking and observes success and failure, respectively. On the other hand, $P_k$ equals $p_k$ weighted by the prior, and thus roughly captures the informativeness generated by effort. As in the example above, we expect skill and effort to be complements when the effort effect $P$ (analogous to $p_k$ in the example) ``increases in $k$ faster'' than the shirking effects $Q^S$ and $Q^F$ (analogous to $p_{0k}$ in the example).

Accordingly, we say that the contracting environment demonstrates \emph{skill-effort complementarity} if $P$ first-order stochastically dominates both $Q^S$ and $Q^F$
\[\sum_{j=1}^k P_j\leq\sum_{j=1}^k Q^S_j\text{ and }\sum_{j=1}^k P_j\leq\sum_{j=1}^k Q^F_j\text{ for all $k$, and some inequality is strict.}\]
The following result shows that this condition is sufficient for strategic uncertainty.
\begin{proposition}\label{claim 1}
    Suppose the post-employment value is strictly increasing in first-order stochastic dominance. Then, strategic uncertainty is present and $F^*$ exhibits contract dispersion if the contracting environment demonstrates skill-effort complementarity.
\end{proposition}

Skill-effort complementarity implies (\ref{criterion for strategic uncertainty}) because, under it, the worker's effort makes success [failure] ``better'' [``worse''] news for the market. In fact, we show in the proof that, with $P$ dominating $Q^S$ [$Q^F$], a market observing success [failure] holds a more optimistic [pessimistic] belief given greater effort.

Skill-effort complementarity fails when $P$ is weakly first-order stochastically dominated by both $Q^S$ and $Q^F$. A special case arises when $p_{0k}$ is ability-independent and $p_k$ is decreasing in the ability ranking induced by $v(\cdot)$. Moreover, when $p_k$ is also ability-independent, which implies $P_k=Q_k^S=Q_k^F$ for all $k$, career value does not vary with conjectured worker effort. This latter case resembles the classic career-concerns models (e.g., \cite{holm1999}; \cite{gimu1992}), which consider additive monitoring systems whose informativeness about ability is independent of effort. We examine this point further in later paragraphs on Classic Monitoring.

\paragraph{Skill-Wage Assortativeness.} How often does strategic uncertainty arise? This subsection focuses on a linear-value setup, which lets us represent the necessary and sufficient condition (\ref{criterion for strategic uncertainty}) through a simple summary statistic and connect our result to empirical evidence, answering the question above.

Consider the following: the post-employment value takes a linear form, namely, $v(\mu)=\sum_{k\in K}u_k\mu_k$ with each ability level $k$'s market value $u_k\geq0$. Moreover, the shirking success probability $p_{0k}$ is ability-independent. We will later refer to this setup as the \emph{linear environment} when we study comparative statics. In this example, the career value function depends on market values $u=(u_k)_{k\in K}$ and effort productivity $p=(p_k)_{k\in K}$ as follows
\begin{equation}\label{career value-linear environment}
    D(q)=\Cov(u,p)T(q;\E_0[p_k]),
\end{equation}
where $\Cov(u,p)$ is the covariance between market values $u$ and effort productivity $p$
\begin{gather*}
    \Cov(u,p)=\E_0[u_kp_k]-\E_0[u_k]\E_0[p_k].
\end{gather*}
The function $T$ is strictly positive and increases strictly in the worker's total working probability $q$ when $q>0$, and $p$ enters $T$ only through its expectation.\footnote{In fact, $T(q;\E_0[p_k])=q/[(p_0+\E_0[p_k]q)(1-p_0-\E_0[p_k]q)]$ where $p_0=p_{0k}$ is the ability-independent shirking success probability.} Note that $\Cov(u,p)=\Corr(u,p)\Std(u)\Std(p)$ where the correlation term $\Corr(u,p)$ determines the sign of $\Cov(u,p)$. Therefore, by assuming nontrivial variation in market value $u$ and productivity $p$, that is, $\Std(u)>0$ and $\Std(p)>0$, we identify a summary statistic for determining strategic uncertainty.
\begin{proposition}\label{claim 2}
    In the linear environment, strategic uncertainty is present and $F^*$ exhibits contract dispersion if and only if $\Corr(u,p)>0$.
\end{proposition}

As a result, employers face a robustness concern if and only if worker skills $p_k$ and their pay in the labor market $u_k$ are matched assortatively. This condition very likely holds in practice, as extensive evidence documents systematic sorting of high-skill workers into high-wage positions (e.g., \cite{abkm1999}; \cite{cahk2013}; \cite{sopg2019}, among many others). Therefore, our result suggests that strategic uncertainty is a prevalent concern in contract design.

\paragraph{Classic Monitoring.} The classic career-concerns settings (e.g., \cite{holm1999}; \cite{gimu1992}) adopt an additive monitoring system, $y=a+k+\epsilon$, where $a$ is the worker's action, $k$ his ability, and $\epsilon$ independent noise. There, the equilibrium distribution of ability and market posteriors does not vary with effort, so career concerns are pinned down by the post-employment value alone and no multiplicity arises. Appendix \ref{classic monitoring system} formalizes this within a framework that generalizes \cite{gimu1992}: career concerns enter the equilibrium condition only as a constant, so criterion (\ref{criterion for strategic uncertainty}) cannot hold and strategic uncertainty is absent.

Multiplicity instead requires effort-dependent informativeness, for example under the multiplicative monitoring of \cite{dejt1999b}. Our framework accommodates general monitoring systems, pins down a sharp criterion for strategic uncertainty, and characterizes optimal contracting.

\subsection{Implications for Contract Dispersion}\label{implications}

In this section, our comparative statics establish that greater contract dispersion occurs when career concerns play a more important role in incentive provision, and when labor markets exhibit greater skill-wage assortativeness, larger skill premiums, and higher occupational mobility. Our results appear surprising because they explain how these variables can shape the \emph{residual} contract dispersion that worker heterogeneity cannot account for. This contribution stems from our focus on ex-ante identical workers who are uninformed about their own characteristics (ability levels), so that a non-degenerate $F^*$ indicates the endogenous use of within-group contract dispersion.

Throughout our analysis, we use superscripts to distinguish parameter sets and their corresponding robustly optimal incentive policies. For instance, we let $v^1$ and $v^2$ denote two post-employment values, while $F^{*1}$ and $F^{*2}$ are the associated robustly optimal incentive policies. Unless otherwise specified, any parameters not mentioned in a given result are assumed identical across the two settings.

To compare the robustly optimal incentive policies under different parameters, we consider a notion of \emph{more dispersed} that simultaneously delivers comparisons across many common inequality measures. The standard statistics used in labor-inequality studies (pay range, standard deviation, mean absolute deviation, top- and bottom-share gaps, the Theil index, and so on) are all expectations of convex functions of the de-meaned (centered) pay. We thus compare the de-meaned distributions in the convex order

\begin{definition}\label{definition-more dispersed}
    \emph{Take two incentive policies $F^1$ and $F^2$ with interval supports $\supp(F^1)=[\underline{b}^1,\overline{b}^1]$ and $\supp(F^2)=[\underline{b}^2,\overline{b}^2]$. $F^2$ is} more dispersed \emph{than $F^1$ if, for every convex function $\phi:\mathbb{R}\to\mathbb{R}$,
    \begin{equation*}
        \E_{F^1}[\phi(b-\E_{F^1}[b])]\;\le\;\E_{F^2}[\phi(b-\E_{F^2}[b])].
    \end{equation*}}
\end{definition}

Definition \ref{definition-more dispersed} implies, in particular, that $F^2$ has a wider support and a larger standard deviation than $F^1$.\footnote{Taking $\phi(x)=x^2$ in the definition yields $\Std(F^1)\le\Std(F^2)$. With $L_1:=\overline{b}^2-\E_{F^2}[b]$ and $L_2:=\underline{b}^2-\E_{F^2}[b]$, taking $\phi(x)=\max\{x-L_1,0\}$ and its mirror $\phi(x)=\max\{L_2-x,0\}$ shows that the centered support $[L_2,L_1]$ of $F^2$ contains that of $F^1$, $[\underline{b}^1-\E_{F^1}[b],\overline{b}^1-\E_{F^1}[b]]$, and thus $\overline{b}^2-\underline{b}^2\ge\overline{b}^1-\underline{b}^1$.} The definition therefore aligns our theoretical comparison with empirical evidence, where contract dispersion is typically measured by the range or the standard deviation of pay.

The following result holds in the general setting and considers discounting the continuation value that workers expect in the labor market. In other words, it varies the share of implicit career-concerns incentives in the worker's total work incentives.

\begin{proposition}\label{proposition-occupational mobility}
    Take two post-employment values $v^1$ and $v^2$. Suppose $F^{*2}$ exhibits contract dispersion. Then, $F^{*2}$ is more dispersed than $F^{*1}$ if there is a discount factor $\delta\in[0,1)$ such that $v^1=\delta v^2$.
\end{proposition}

Proposition \ref{proposition-occupational mobility} establishes that when workers place more weight on future income, employers tend to expand contract dispersion. What drives the result is the \emph{share} of total incentives that career concerns supply. The discount factor $\delta$ scales the implicit incentive $D(\cdot)$ relative to the explicit bonus, and with it the gap $\overline{b}-\underline{b}$ between the partial-implementation and deterministic robust bonuses. Since this gap is precisely the room that strategic uncertainty opens for dispersion, implicit incentives looming larger in motivating workers (due, for example, to greater patience, stronger protection for job changes, or higher labor mobility) render explicit incentives more dispersed, as employers optimally respond to their robustness concern. The comparative static thus operates through the composition of incentives rather than their level, linking incentive structure directly to inequality.

One concrete source of variation in $\delta$ is occupational mobility. Workers discount by $\delta=e^{-rt}$, where $r$ is the discount rate and $t$ the combined duration of current employment and subsequent job search, so higher mobility across firms reduces $t$, raises $\delta$, and amplifies the role of career concerns. This positive association between mobility and inequality is well documented (e.g., \cite{blmi2002} and \cite{kama2009}).

Furthermore, our second comparative statics result focuses on the linear-value example discussed in Section \ref{interpreting the criterion}, where the post-employment value is linear, $v(\mu)=\sum_{k\in K}u_k\mu_k$, and the shirking success probabilities are ability-independent, $p_{0k}=p_0$. We have referred to this setting as the linear environment. Denote by $u=(u_k)_{k\in K}$ the profile of market values and by $p=(p_k)_{k\in K}$ the profile of productivity.

\begin{proposition}\label{proposition-assortativeness skill premium}
    Consider the linear environment. Take two market values $u^1$ and $u^2$, and two productivities $p^1$ and $p^2$. Suppose $F^{*2}$ exhibits contract dispersion, and maintain $\E_0[p^1_k]=\E_0[p^2_k]$. Then, $F^{*2}$ is more dispersed than $F^{*1}$ if $\Cov(u^2,p^2)>\Cov(u^1,p^1)$. Moreover, this condition holds if $p:=p^1=p^2$, $\E_0[u^1_k]=\E_0[u^2_k]$, and $u^2$ single-crosses $u^1$ from below, that is, there is $k^*\in K$ such that $u^2_k>[<]u^1_k$ for all $p_k>[<]p_{k^*}$.
\end{proposition}

Proposition \ref{proposition-assortativeness skill premium} identifies two determinants of residual pay inequality. The first part of the result predicts that greater dispersion emerges when more productive workers (whose effort raises the success probability more) are more likely to match with higher-paying positions. As an example, consider $ K=\{1,2,...,K\}$ with uniform prior. We let market 2 have ``efficient matching'' with both $u^2_k$ and $p^2_k$ strictly increasing in $k$, yet we introduce some ``mismatch'' in market 1 by setting $u^1_k=u^2_{t(k)}$ where $t$ is a permutation of $K$. We hold $p^1=p^2$ and $\Std(u^1)=\Std(u^2)$. Hence, we have $\Corr(u^2,p^2)>\Corr(u^1,p^1)$ and $\Cov(u^2,p^2)>\Cov(u^1,p^1)$. Proposition \ref{proposition-assortativeness skill premium} then implies that market 2 has greater contract dispersion. This prediction is in line with \cite{cahk2013}, who show that the increasing assortativeness in the assignment of workers to plants contributes to the expanding dispersion of West German wages. Our result provides a novel mechanism underlying their findings: assortative skills $p$ and wages $u$ make effort and career concerns complementary, leading to strategic uncertainty in contracting; employers optimally respond by creating inequality.

The second part of Proposition \ref{proposition-assortativeness skill premium} holds skills fixed at $p^1=p^2$ while varying market value in a single-crossing direction, with rising skill premiums as a special case. In particular, consider $ K=\{1,2,...,K\}$ with $p_k$ strictly increasing in $k$. Construct $u^2$ to be steeper than $u^1$, namely $u^2_k-u^2_{k-1}>u^1_k-u^1_{k-1}$ for all $k>1$. By normalizing $\E_0[u^1_k]=\E_0[u^2_k]$, we ensure that $u^2$ single-crosses $u^1$ from below. The parallel rise in skill premiums and inequality has been well documented, for instance by studies of skill-biased technical change and pay inequality (see \cite{korv2000}, \cite{acem2002}, and the citations therein). While the standard explanation attributes this relationship to pay differentiation across skill levels, we identify a novel channel: with greater skill premiums in the labor market, career concerns become more important relative to explicit contract incentives, and thus strategic uncertainty looms larger. What is surprising is that skill premiums shape inequality even after we remove worker heterogeneity (and hence any scope for cross-group discrimination).

By decomposing covariance as $\Cov(u,p)=\Corr(u,p)\Std(u)\Std(p)$, we see that contract dispersion increases when we strengthen the relationship between market value $u$ and skill $p$ in two distinct ways: we can either increase their correlation $\Corr(u,p)$ (assortativeness) or expand the spread of market value $\Std(u)$ (skill premiums).

\section{Extensions}\label{discussions}

\subsection{Contractual Transparency}\label{contractual transparency}

The main model assumes that the labor market cannot observe the worker's realized contract. We now examine this assumption: we contrast our private-contracting setting with a public-contracting benchmark, link the assumption to the legal and regulatory environment that justifies it, and study how the analysis changes under partial leakage of contract information.

\paragraph{Public-contracting benchmark.} Consider the alternative environment in which the market observes not only the incentive policy $F$ but also the realized contract $b$. The employer's optimal incentive policy then collapses to the deterministic robust bonus, a degenerate distribution at $\overline{b}$. Under public contracting, each contract realization (rather than the distribution) induces a worker-market subgame, and the employer minimizes her expected cost independently in each subgame. It is therefore without loss to offer a single contract that fully implements work at the least cost. The benchmark delivers a sharp lesson on the role of privacy in our main model: randomization is useful \emph{only} because of the joint presence of contractual privacy and the robustness concern. Neither force alone leads to contract dispersion.

\paragraph{Implications for the privacy debate.} Our private-contracting assumption reflects the current legal and regulatory environment that restricts firms' access to workers' compensation histories.\footnote{For example, 22 states and 24 local districts in the US have enacted \href{https://www.hrdive.com/news/salary-history-ban-states-list/516662/}{Salary History Bans} since 2017, which forbid future employers from inquiring about workers' past contracts. The Supreme Court case \href{https://law.justia.com/cases/federal/appellate-courts/ca9/16-15372/16-15372-2020-02-27.html}{Rizo v.\ Yovino 2020} determined that contracting should not perpetuate past pay inequality. See, for example, \cite{bedm2024} for a recent survey on the debate about the impact of these privacy practices.} These policies have generated an active debate about their effect on pay inequality. A prominent view (e.g., \cite{sinh2019}; \cite{hamc2020}; \cite{bedm2024}) holds that contractual-privacy policies such as Salary History Bans reduce pay disparities across observable groups such as gender and race. Our model identifies a within-group margin that this debate has overlooked: firms' motives to provide incentives and rule out low-effort beliefs generate contract dispersion among observationally identical workers if and only if contracting is private. Consequently, contractual-privacy policies may inadvertently shift inequality from across groups to within groups, potentially overstating their net effect.\footnote{Which channel dominates remains an open empirical question. For instance, \cite{srvw2020} and \cite{daow2022} find that cross-group pay convergence is often negligible, while others report only modest (e.g., 1\% in \cite{hamc2020}) or mixed (e.g., \cite{sinh2019}) effects. Our result thus calls for a re-evaluation of these policies.}$^,$\footnote{The same evidence also documents that Salary History Bans reduce average pay (e.g., \cite{srvw2020}; \cite{daow2022}). Our prediction is consistent with this because, compared to public contracting, the employer with robustness concerns manages to lower contract offers by shifting market expectations. In contrast, when the employer ignores strategic uncertainty and merely implements work partially, her contract choice is the same with or without contractual privacy.}

\paragraph{Partial leakage of contract information.} The main model and the public-contracting benchmark are two extreme cases. In practice, the privacy policies above are unlikely to be perfectly enforced. We now consider four intermediate scenarios in which contracting is neither completely private nor completely public: (i) the market observes the contract with some probability; (ii) the worker can disclose the contract with some probability; (iii) the employer can release information about the contract; or (iv) some firms in the market can observe the contract.

The main message of the paper will not change in any of these four cases because strategic uncertainty depends on the same criterion (\ref{criterion for strategic uncertainty}). To see this, consider restricting the employer to a single contract rather than a distribution, so that the policy is deterministic and communication about the contract is irrelevant. Section \ref{strategic uncertainty} has demonstrated that the partial-implementation and deterministic robust bonuses here are $\underline{b}$ and $\overline{b}$, respectively, and strategic uncertainty emerges if and only if they diverge.

To understand why randomization remains helpful in cases (i) and (ii), we assume that the ``silent'' event (in which no contract information can be transmitted) occurs with positive probability. Hence, randomization can still shift market expectations by exploiting the market's ignorance in the silent event. The departure from our main result is that, in iteratively constructing the robustly optimal incentive policy, the employer must guarantee work at each contract while accounting for the equilibrium behavior in the non-silent event. For case (iii), randomization improves upon the deterministic robust bonus when, for example, the employer commits not to release any information and uses the robustly optimal incentive policy from Theorem \ref{main result}. To understand how randomization emerges in case (iv), we consider the following setting. Denote by $\mu_O$ the posterior belief of the firms that observe the contract, and by $\mu_N$ that of the firms that do not. The worker's post-employment value is the highest contract offer from the market, $v(\mu_O,\mu_N):=\max\{v_0(\mu_O),v_0(\mu_N)\}$, where $v_0(\mu)$ is a firm's offer when it holds belief $\mu$. Recall that the main goal in guaranteeing effort is to lure the worker to work at bonuses where he currently shirks. When these bonuses are realized, the firms that observe the contract know that the worker shirks ($q=0$). Therefore, the career value relevant for inducing this deviation is given by
\[D(q)=\max\Big\{v_0(\overline{\mu}(q)),v_0(\overline{\mu}(0))\Big\}-\max\left\{v_0(\underline{\mu}(q)),v_0(\underline{\mu}(0))\right\}.\]
To convey the intuition, suppose that we have skill-effort complementarity. In this case, the uninformed firms have more reactive beliefs, implying $v_0(\overline{\mu}(q))\geq v_0(\overline{\mu}(0))$ and $v_0(\underline{\mu}(q))\leq v_0(\underline{\mu}(0))$. In other words, the worker joins the informed firms if he fails and the uninformed firms if he succeeds. As a result, the modified career value $D(q)=v_0(\overline{\mu}(q))-v_0(\underline{\mu}(0))$ is attenuated relative to the standard definition (\ref{career value}), but it remains increasing because $v_0(\overline{\mu}(\cdot))$ is increasing under skill-effort complementarity. The optimal contract structure will exhibit less dispersion, but dispersion will still be used to mitigate strategic uncertainty.

\subsection{Informed Worker}\label{informed worker}

The main model assumes that the worker does not know his ability. This section relaxes this assumption: the worker observes his ability $k\in K$ privately, so $k$ is his \emph{type}. He thus knows his shirking success probability $p_{0k}$ and his additional success probability from working $p_k$. The working cost is again $c$. We again ask: what is the robustly optimal incentive policy? The timing is unchanged, except that the worker observes his ability before the work decision: given incentive policy $F$, upon receiving a contract with bonus $b$, each worker type $k$ chooses a working probability $\sigma_k(b)$. The market, observing only the incentive policy and the project outcome, forms a posterior belief about the worker's ability and pays him the post-employment value.

We begin by defining career value for the new specification, then describe the equilibrium and the assumption we use. Each worker type's work decision $\sigma_k$ induces a type-contingent total working probability $q_k:=\E_F[\sigma_k(b)]=\int_{\mathbb{R}_+}\sigma_k(b)dF(b)$. The market's equilibrium beliefs contingent on success and failure are $\overline{\mu}$ and $\underline{\mu}$, respectively. They now depend on the type-contingent total working probabilities $(q_k)_{k\in K}$ through Bayes' rule. Thus, we write them as $\overline{\mu}((q_k)_{k\in K})$ and $\underline{\mu}((q_k)_{k\in K})$, which are again continuous functions. The career value is then defined accordingly
\begin{equation*}
    D((q_k)_{k\in K})=v(\overline{\mu}((q_k)_{k\in K}))-v(\underline{\mu}((q_k)_{k\in K})).
\end{equation*}
The worker strategy profile $(\sigma_k)_{k\in K}$ forms a PBE if, for every type $k$ and bonus realization $b$
\begin{equation}\label{equilibrium concept-private type}
    \sigma_k(b)\in\argmax_{\widetilde{\sigma}\in[0,1]}\widetilde{\sigma}[b+D((q_k)_{k\in K})-\lambda_k]\text{, where }q_k=\E_F[\sigma_k(b)],
\end{equation}
where $\lambda_k:=\frac{c}{p_k}$ is type $k$'s effective cost. Different worker types face heterogeneous work incentives, captured by $b+D-\lambda_k$ in (\ref{equilibrium concept-private type}), because the effective cost varies with worker ability. More specifically, in an equilibrium with career value equal to $D$, the type-$k$ worker's threshold bonus is $b_k:=\lambda_k-D$, which pins down the worker's strategy (except for his randomization decision at $b_k$).

To keep the section compact, we focus on binary types $ K=\{H,L\}$ with the high [low] type having a low [high] effective cost, namely $\lambda_H<\lambda_L$. We first characterize the robustly optimal incentive policy for implementing full working. In Appendix \ref{private types: partial working and multiple types}, we extend the result to cases with either partial working or multiple types.

Our result here depends on the following assumption: the post-employment value $v(\cdot)$ is continuous and strictly increasing in the high type's probability. This means ability and market pay are assortative. To see this, $\lambda_H=\frac{c}{p_H}<\frac{c}{p_L}=\lambda_L$ means that the high type is more skilled, in the sense that he raises the success probability more effectively. Therefore, $v(\cdot)$ increasing in high-type probability captures a market that values skill. Under this assumption, the career value function $D(q_L,q_H)$ is strictly increasing in $q_H$ and decreasing in $q_L$: the high [low] type's work raises success probability more [less] effectively, which renders success a stronger [weaker] signal of high ability, widening [narrowing] the continuation-value gap. We thus consider an environment in which the two types have asymmetric externalities, that is, the high [low] type's work increases [decreases] work incentives.

\paragraph{Analysis.} The analysis likewise begins with two critical bonuses
\begin{equation*}
    \underaccent{\sim}{b}=\lambda_L-D(1,1);\quad\quad\quad\accentset{\sim}{b}=\max\{\lambda_H-D(0,0),\underaccent{\sim}{b}\}.
\end{equation*}
Notice that $\underaccent{\sim}{b}$ is the partial-implementation bonus because in a full-working equilibrium, the career value is $D(1,1)$, and $\underaccent{\sim}{b}$ induces work from the worker with lower work incentives. Another bonus $\lambda_H-D(0,0)$ (in the definition of $\accentset{\sim}{b}$) is the lowest bonus needed to break the bad equilibrium in which no working happens, where the career value is $D(0,0)$, and this bonus lures work from the worker with higher work incentives. In fact, we show (see Proposition \ref{proposition-criterion for contract dispersion-general} in Appendix \ref{private types: partial working and multiple types}) that $\accentset{\sim}{b}$ is the minimal bonus for fully implementing full working (i.e., the deterministic robust bonus), and that the partial-implementation bonus $\underaccent{\sim}{b}$ suffers from multiplicity if and only if the following holds
\begin{equation}\label{criterion for strategic uncertainty-private type}
    \accentset{\sim}{b}>\underaccent{\sim}{b}\text{, or equivalently, }D(1,1)-D(0,0)>\lambda_L-\lambda_H.
\end{equation}
This extends criterion (\ref{criterion for strategic uncertainty}). The difference is that, with the worker informed, his effort must separate the market's post-success and post-failure beliefs by a sufficiently large margin, since $\lambda_L-\lambda_H>0$.

Proposition \ref{main result-private type} states the extension result: the robustly optimal incentive policy $F^{\dagger}$ is fully supported on $[\underaccent{\sim}{b},\accentset{\sim}{b}]$, with a mass point at the minimal bonus and otherwise continuous. Hence, within-skill contract dispersion arises under the same criterion (\ref{criterion for strategic uncertainty-private type}). To maintain compactness, we relegate the formal analysis and statement to Appendix \ref{private types: partial working and multiple types}.

\begin{proposition}[Informal]\label{main result-private type}
    To induce full working from a binary-type informed worker, the unique robustly optimal incentive policy $F^{\dagger}$ has $\supp(F^{\dagger})=[\underaccent{\sim}{b},\accentset{\sim}{b}]$, is continuous on $(\underaccent{\sim}{b},\accentset{\sim}{b}]$, and has a mass point at $\underaccent{\sim}{b}$.
\end{proposition}

To understand Proposition \ref{main result-private type}, consider the following procedure that pins down the robustly optimal incentive policy. First, the employer offers the bonus $\accentset{\sim}{b}$ that, given skeptical market expectations and thus small career concerns, attracts only the high type into work while discouraging the low type. This guarantees larger career concerns, since the market now attributes project success [failure] to high-type working [low-type shirking]. As a result, the employer can ensure greater effort with lower bonuses. The employer then sequentially assigns probability to successively lower bonuses, starting from $\accentset{\sim}{b}$, to keep strengthening career concerns by maintaining that the high type works ``more than'' the low type. Eventually, the procedure ends once it reaches the equilibrium-keeping bonus $\underaccent{\sim}{b}$, leaving a mass point at the bottom.

\paragraph{Generalization.} To extend the result beyond the binary-type, full-working case, we consider partial working and multiple types in Appendix \ref{private types: partial working and multiple types}. To robustly induce a given profile of working probabilities, the employer may introduce multiple mass points (see Proposition \ref{main result-Q}). With multiple types,
Proposition \ref{proposition-criterion for contract dispersion-general} extends criterion (\ref{criterion for strategic uncertainty-private type}) and shows that our main message is preserved in general: dispersion arises precisely when strategic uncertainty is present, which typically reflects complementary reputation effects in labor markets.

\subsection{Career Concerns through Two-Period Contracting}\label{Career Concerns through Two-Period Contracting}

This section micro-founds the source of career concerns. We replace the exogenous post-employment value $v(\cdot)$ in the main model with a second-period employer who optimally chooses an incentive policy. We show that the first-period employer's problem coincides with that of the main model, with the career value $D(\cdot)$ now micro-founded by the second-period employer's contingent offers. We also give a numerical example in which $D(\cdot)$ is strictly increasing, illustrating the complementarity regime that generates contract dispersion in the optimal first-period policy.

The two periods, indexed by $t\in\{1,2\}$, are run by distinct employers $E_1$ and $E_2$ on one-shot projects with the same primitives as in the main model. The worker's ability is persistent, drawn from $\mu^0$, and project outcomes are independent across periods conditional on ability. Unlike the main model, there is no labor market after period 2. A second departure from the main model is that each employer's target effort is endogenous: both employers share a strictly concave value function $u:[0,1]\to\mathbb{R}_+$ over the working probability $q_t$ of their own project at time $t$. This can be viewed as a valuation of the success probability, which is pinned down by the working probability. We assume $u$ is such that the optimal $q_t$ is interior (e.g., $u'(0)=\infty$, $u'(1)=0$). $E_t$'s payoff is $u(q_t)$ net of her expected period-$t$ bonus payment. The main model abstracts from this margin because implementing any working probability is analogous to implementing full work. Here, this remains the case for $E_1$'s problem, but $E_2$'s optimal effort level can vary with the period-1 outcome, a feature we preserve. The period-1 timing replicates the main model: $E_1$ publicly offers a policy $F_1\in\Delta(\mathbb{R}_+)$, the worker privately observes his bonus $b_1$ and chooses his effort $a_1$, and the outcome $x_1\in\{0,1\}$ is publicly realized. In period 2, $E_2$ observes $(F_1,x_1)$ but not $(b_1,a_1)$, and publicly offers a period-2 policy $F_2\in\Delta(\mathbb{R}_+)$. The worker privately observes $b_2$ drawn from $F_2$, chooses $a_2$, and the second outcome $x_2$ is realized.

To specify $E_1$'s problem, we define the set of equilibria $\mathcal{E}(F_1)$ induced by each period-1 policy $F_1$. An equilibrium of the continuation game is a tuple $g=(\sigma_1,\sigma_2,\phi,\overline{\mu},\underline{\mu},\nu)$, where $\sigma_1$ is the worker's period-1 working probability as a function of the bonus realization $b_1$; $\phi$ maps the period-1 outcome to $E_2$'s policy offer; $\sigma_2$ is the worker's period-2 working probability and $\nu$ is his period-2 belief, both as functions of his period-2 information (the period-1 action and outcome, and both periods' bonus realizations); and $(\overline{\mu},\underline{\mu})$ are $E_2$'s posteriors contingent on period-1 success and failure, respectively. The tuple is an equilibrium if (i) $\sigma_1$ and $\sigma_2$ are optimal given the worker's information in their respective periods; (ii) for each $x_1$, $\phi(x_1)$ maximizes $E_2$'s expected payoff given her posterior, namely $\overline{\mu}$ if $x_1=1$ and $\underline{\mu}$ if $x_1=0$;\footnote{The multiplicity issue disappears in the second period because career concerns are absent. With more than two periods, downstream contracting may face robustness concerns, in which case the set of continuation equilibria induced by each period's policy must be defined recursively, starting from the last period: given the worker's strategies in earlier periods, the policies offered so far, and the optimal behavior of later employers, the worker's current-period strategy is optimal. This nested-robustness setting is, however, beyond the scope of the paper.} and (iii) the beliefs satisfy Bayes' rule. We collect all such equilibria in $\mathcal{E}(F_1)$.

$E_1$'s problem mirrors that of the main model: she chooses $F_1$ to maximize her expected payoff, $u(q_1)$ net of her expected period-1 bonus payment, subject to all equilibria inducing the same period-1 working probability $q_1$, that is, $q_1=\E_{F_1}[\sigma_1(b_1)]$ for all $g\in\mathcal{E}(F_1)$.

\paragraph{Analysis.} We show below that the worker's period-1 strategy reacts to a career value $D(q_1)$, a function of the expected period-1 working probability $q_1$. Hence, $E_1$'s robust implementation of any $q_1$ coincides with that in the main model.

\emph{Private beliefs.} The worker's period-2 belief, derived by Bayes' rule from his realized period-1 action $a_1$ and the observed outcome $x_1$ (with $x_1=1$ indicating success), takes one of four values
\begin{equation*}
    \nu_{a_1,x_1}(k)\propto\mu^0_k(p_{0k}+a_1p_k)^{x_1}(1-p_{0k}-a_1p_k)^{1-x_1},\qquad (a_1,x_1)\in\{0,1\}^2,
\end{equation*}
and the corresponding private effective costs are
\begin{equation*}
    \lambda_{a_1,x_1}:=c/\E_{\nu_{a_1,x_1}}[p_k].
\end{equation*}
For each $x_1$, let $\overline\lambda_{x_1}:=\max_{a_1}\lambda_{a_1,x_1}$ denote the higher of the two private effective costs at that outcome. Note that $\nu_{a_1,x_1}$, $\lambda_{a_1,x_1}$, and $\overline{\lambda}_{x_1}$ are pinned down by model primitives.

\emph{$E_2$'s problem.} Fix the conjectured $q_1$ and the observed $x_1$. $E_2$'s posterior $\mu(x_1;q_1)\in\{\overline\mu(q_1),\underline\mu(q_1)\}$ is the Bayesian mixture of $\nu_{0,x_1}$ and $\nu_{1,x_1}$ weighted by the conditional probabilities of $a_1$ given $(x_1,q_1)$, and depends on $q_1$ only through these weights. To induce work from both worker types in the work-eliciting branch of her policy, $E_2$'s cost-minimizing instrument is the two-point policy (where $\delta_b$ denotes a degenerate contract with bonus $b$)
\begin{equation}\label{E2 policy}
    F_2(x_1;q_1)=q_2(x_1;q_1)\,\delta_{\overline\lambda_{x_1}}+\bigl(1-q_2(x_1;q_1)\bigr)\,\delta_0,
\end{equation}
under which the high realization $\overline\lambda_{x_1}$ leaves the higher-effective-cost worker just willing to work and the lower-effective-cost one strictly willing to work, while the zero realization induces shirking from both. The induced period-2 working probability is $q_2$, and the expected period-2 payment is $q_2\overline\lambda_{x_1}\E_{\mu(x_1;q_1)}[p_{0k}+p_k]$. $E_2$'s problem hence reduces to
\begin{equation*}
    \max_{q_2\in[0,1]}u(q_2)-q_2\overline\lambda_{x_1}\E_{\mu(x_1;q_1)}[p_{0k}+p_k],
\end{equation*}
whose unique interior solution $q_2^*(x_1;q_1)$ satisfies the first-order condition
\begin{equation}\label{E2 FOC}
    u'\bigl(q_2^*(x_1;q_1)\bigr)=\overline\lambda_{x_1}\E_{\mu(x_1;q_1)}[p_{0k}+p_k].
\end{equation}
Since $\overline\lambda_{x_1}$ does not depend on $q_1$, $q_2^*$ depends on $q_1$ only through $E_2$'s posterior belief $\mu(x_1;q_1)$.

\emph{Worker's period-2 utility.} Combining \eqref{E2 policy} with the worker's private belief, we find that his expected period-2 utility decomposes simply: the zero realization yields zero, and the high realization $\overline\lambda_{x_1}$ yields his net utility,
\begin{equation*}
    V(a_1,x_1):=\overline\lambda_{x_1}\,\E_{\nu_{a_1,x_1}}[p_{0k}+p_k]-c.
\end{equation*}
Note that $V(a_1,x_1)$ is also pinned down by model primitives.

\emph{Career value.} The worker receiving $b_1$ compares his expected total payoff from $a_1=1$ and $a_1=0$, weighting the continuation payoff by the ability-conditional outcome probability $\Pr(x_1\mid k,a_1)=(p_{0k}+a_1p_k)^{x_1}(1-p_{0k}-a_1p_k)^{1-x_1}$. Let $\Pr(x_1\mid \mu,a_1):=\sum_{k\in K}\mu_k\Pr(x_1\mid k,a_1)$ denote the outcome probability given belief $\mu$. As a result, the period-1 incentive constraint collapses to a familiar rule
\begin{equation*}
    b_1+D(q_1)\geq\lambda,
\end{equation*}
where $\lambda=c/\E_{0}[p_k]$ is the main-model effective cost, whereas the career value becomes
\begin{equation*}
    D(q_1):=\frac{1}{\E_{0}[p_k]}\sum_{x_1\in\{0,1\}}U_{x_1}q_2^*(x_1;q_1) \text{, where }U_{x_1}:=\Pr(x_1\mid\mu^0,1)\,V(1,x_1)-\Pr(x_1\mid\mu^0,0)\,V(0,x_1).
\end{equation*}
The career value is a linear combination of the two period-2 working probabilities $q_2^*(1;q_1)$ and $q_2^*(0;q_1)$, whose coefficients $U_{x_1}$ are given by model primitives. The $q_1$-dependence of $D(\cdot)$ therefore inherits the smoothness of $q_2^*(x_1;q_1)$, which by the implicit function theorem applied to \eqref{E2 FOC} is a smooth function of $q_1$ through $E_2$'s posterior.\footnote{For a numerical example, consider binary ability $K=\{A,B\}$ with prior $\mu^0_A=\mu^0_B=0.5$, success probabilities $p_A=0.5$, $p_B=0.2$, $p_{0A}=0.1$, and $p_{0B}=0.5$, working cost $c=0.1$, and employer valuation $u(q)=0.3\sqrt{q}$. Note that $u'(1)=0$ fails, but what matters is to have interior optimal $q$. The resulting career value is a strictly increasing function.}

\paragraph{Equivalence with the main model.} The worker in this setting still trades off the explicit incentive $b_1$ against the career value $D(q_1)$, and $E_1$'s robust implementation of any target $q_1$ runs through the same machinery as our main characterization. The substantive change is the micro-foundation of $D(\cdot)$: rather than being borrowed from a black-box labor-market valuation, $D(q_1)$ now incorporates the second employer's value-cost trade-offs at the two on-path posteriors, as (\ref{E2 FOC}) makes explicit.

\bibliographystyle{aer}
\bibliography{bib}

\appendix
\renewcommand{\thesection}{Appendix \Alph{section}}
\renewcommand{\thesubsection}{\Alph{section}.\arabic{subsection}}

\section{}

\subsection{Proof of Proposition \ref{proposition-benchmarks}}

First, we show that the partial-implementation policy, denoted as $F^P$, assigns all the probability to $\underline{b}$. This is true because, in the full-working equilibrium, the worker faces a career value $D(1)$. Thus, this equilibrium exists if and only if an incentive policy does not assign any positive probability to bonuses lower than $\lambda-D(1)$, equivalent to $\supp(F^P)\subset[\underline{b},\infty)$. Moreover, any such incentive policy can be improved by shifting all the probability to $\underline{b}$, which lowers the expected cost while ensuring the existence of the desired equilibrium. Therefore, the incentive policy degenerate at $\underline{b}$ is the (almost surely) unique solution to the partial-implementation benchmark.

Second, to determine the deterministic robust bonus, denoted as $b^R$, we characterize the equilibrium set induced by each deterministic bonus, say $b$. There are three cases: (i) if $b\geq\lambda-D(1)=\underline{b}$, the worker works with probability one; (ii) if $b\leq\lambda-D(0)$, the worker shirks; (iii) if $b=\lambda-D(q)$, the worker randomizes and works with probability $q$. Therefore, to ensure that case (i) exists while case (ii) does not, we must have $b^R\geq\underline{b}$ and $b^R>\lambda-D(0)$. To make sure that case (iii) vanishes in the equilibrium set, we need $b^R\neq\lambda-D(q)$ for all $q<1$, which means $b^R\notin\{\lambda-D(q):q<1\}$. The continuity of $D(\cdot)$ further implies that $\{\lambda-D(q):q\in[0,1]\}=[\lambda-\max_{q\in[0,1]}D(q),\lambda-\min_{q\in[0,1]}D(q)]$. Combining these three conditions, $b^R$ fully implements full working if and only if $b^R\geq\overline{b}=\lambda-\min_{q\in[0,1]}D(q)$ if $D(q)$ is uniquely minimized by $q=1$, or $b^R>\overline{b}$, otherwise. In either case, the infimum is $\overline{b}$.

Lastly, we show offering $\underline{b}$ is not robustly optimal if and only if $\underline{b}<\overline{b}$. This result is in fact implied by Theorem \ref{main result}, whose proof does not depend on any argument here. Theorem \ref{main result} shows that the robustly optimal incentive policy is (essentially) unique and fully supported on $[\underline{b},\overline{b}]$. Hence, the degenerate policy at $\underline b$ is robustly optimal if and only if $\underline b=\overline b$, i.e., $D(1)=\min_q D(q)$.


\subsection{Proof of Proposition \ref{claim 1} and Proposition \ref{claim 2}}

First, we show Proposition \ref{claim 1} is true. It suffices to show that with skill-effort complementarity, $D(q)$ is strictly increasing in $q$. In fact, we can show that given the working probability $q$, the market's posterior belief contingent on observing success [failure] $\overline{\mu}(q)$ [$\underline{\mu}(q)$], given by Bayes' rule (\ref{bayes rule}), is strictly increasing [decreasing] in the order of first-order stochastic dominance, so the fact that $v(\cdot)$ increases strictly with respect to strict first-order stochastic dominance yields the result. To do this, we want to show that for all $k$, we have $\frac{d}{dq}\sum_{j=1}^k\overline{\mu}_j(q)\leq0$ and $\frac{d}{dq}\sum_{j=1}^k\underline{\mu}_j(q)\geq0$, and at least one inequality is strict
\begin{equation*}
    \begin{aligned}
        \frac{d}{dq}\sum_{j=1}^k\overline{\mu}_j(q)\leq0&\Leftrightarrow\frac{d}{dq}\frac{\sum_{j=1}^k\mu_j^0(p_{0j}+p_jq)}{\sum_{j=1}^K\mu_j^0(p_{0j}+p_jq)}\leq0 \\
        &\Leftrightarrow\left(\sum_{j=1}^k\mu_j^0p_j\right)\left[\sum_{j=1}^K\mu_j^0(p_{0j}+p_jq)\right]\leq\left(\sum_{j=1}^K\mu_j^0p_j\right)\left[\sum_{j=1}^k\mu_j^0(p_{0j}+p_jq)\right] \\
        &\Leftrightarrow\left(\sum_{j=1}^k\mu_j^0p_j\right)\left(\sum_{j=1}^K\mu_j^0p_{0j}\right)\leq\left(\sum_{j=1}^K\mu_j^0p_j\right)\left(\sum_{j=1}^k\mu_j^0p_{0j}\right) \\
        &\Leftrightarrow\sum_{j=1}^kP_j\leq\sum_{j=1}^kQ^S_j;\text{ and } \\
        \frac{d}{dq}\sum_{j=1}^k\underline{\mu}_j(q)\geq0&\Leftrightarrow\text{ with similar steps }\Leftrightarrow\sum_{j=1}^kP_j\leq\sum_{j=1}^kQ^F_j.
    \end{aligned}
\end{equation*}
By skill-effort complementarity, both of the above hold and, for at least one $k$, one inequality is strict.

Next, we show (\ref{career value-linear environment}), the career value function in the linear environment. By plugging in (\ref{bayes rule}), we have
\begin{equation*}
    \begin{aligned}
        D(q)&=\sum_{k\in K}u_k\overline{\mu}_k-\sum_{k\in K}u_k\underline{\mu}_k=\frac{\sum_{k\in K}u_k\mu^0_k(p_0+p_kq)}{\sum_{k\in K}\mu_k^0(p_0+p_kq)}-\frac{\sum_{k\in K}u_k\left[\mu^0_k-\mu^0_k(p_0+p_kq)\right]}{1-\sum_{k\in K}\mu_k^0(p_0+p_kq)} \\
        &=\frac{\E_0[u_k]p_0+\E_0[u_kp_k]q}{p_0+\E_0[p_k]q}-\frac{\E_0[u_k]-\left(\E_0[u_k]p_0+\E_0[u_kp_k]q\right)}{1-\left(p_0+\E_0[p_k]q\right)} \\
        &=\frac{\E_0[u_k]p_0+\E_0[u_kp_k]q-\E_0[u_k]\left(p_0+\E_0[p_k]q\right)}{\left(p_0+\E_0[p_k]q\right)\left[1-\left(p_0+\E_0[p_k]q\right)\right]} \\
        &=\frac{\left(\E_0[u_kp_k]-\E_0[u_k]\E_0[p_k]\right)q}{\left(p_0+\E_0[p_k]q\right)\left(1-p_0-\E_0[p_k]q\right)}=\Cov(u,p)T(q;\E_0[p_k]),
    \end{aligned}
\end{equation*}
where for all $q>0$, $T$ can be rewritten as follows
\begin{equation*}
        T(q;\E_0[p_k])=\frac{1}{p_0(1-p_0)/q-\left(\E_0[p_k]\right)^2q+(1-2p_0)\E_0[p_k]}.
\end{equation*}
One can see that $T(0;\E_0[p_k])=0$ and $T$ is strictly increasing in $q$ for all $q>0$.

\subsection{Proof of Lemma \ref{lemma-strict increasing}}

Take any $\widetilde{b}\in(\underline{b},\overline{b})$ and $\widetilde{q}:=\max\{q':\widetilde{b}=\lambda-D(q')\}$. We suppose for contradiction there is some $q'>\widetilde{q}$ such that $D(q')\leq D(\widetilde{q})$. Note that $\widetilde{b}>\underline{b}$ implies $D(\widetilde{q})<D(1)$. As a result, we have $D(q')\leq D(\widetilde{q})<D(1)$, and the continuity of $D(\cdot)$ guarantees the existence of some $q\in[q',1)$ that satisfies $\widetilde{b}=\lambda-D(q)$, a contradiction to $\widetilde{q}$ being the largest such probability. Consequently, $D(q')>D(\widetilde{q})$ for all $q'>\widetilde{q}$.

\subsection{Proof of Proposition \ref{proposition-auxiliary problem} and Theorem \ref{main result}}

We first show that every robustly optimal incentive policy must be feasible in the auxiliary problem (\ref{auxiliary problem}). Next, we show that problem (\ref{auxiliary problem}) has a unique solution given by Theorem \ref{main result}. To conclude, we show that the solution can be approximated by a sequence of incentive policies in the sense of Definition \ref{definition-solution concept}, and that it is the unique robustly optimal incentive policy. In this proof, we assume $\overline{b}>\underline{b}$ since the opposite case with $\overline{b}=\underline{b}$ is trivial.\footnote{In this case, offering only bonuses above $\overline{b}$ ensures full implementation of full working, while offering any bonuses below $\underline{b}$ prevents full working from being an equilibrium. Thus, the robustly optimal incentive policy is degenerate at $\overline{b}=\underline{b}$.} Also, define the inverse of a distribution as a set function
\begin{equation*}
    F^{-1}(1-q):=\{b:q\in[1-F(b^+),1-F(b^-)]\}.
\end{equation*}

\emph{Step 1}. We take any robustly optimal incentive policy $\widetilde{F}^*$ and a sequence of incentive policies $(F_n)_{n=1}^{\infty}$ that approximates it in the sense of Definition \ref{definition-solution concept}. Since every $F_n$ fully implements full working, it must satisfy (\ref{equilibrium keeping}) to ensure that there is a full-working equilibrium. Suppose there is $\widehat{b}<\underline{b}$ such that $\widetilde{F}^*(\widehat{b})>0$. Then, the right-continuous nondecreasing function $\widetilde{F}^*$ must be continuous at some $b'\in(\widehat{b},\underline{b})$ with $\widetilde{F}^*(b')>0$. However, this is impossible because $F_n(b')=0$ for all $n\geq1$ and weak convergence implies $F_n(b')\rightarrow \widetilde{F}^*(b')$. Hence, $\widetilde{F}^*$ also satisfies (\ref{equilibrium keeping}).

Moreover, for all $q\in[0,1)$ and $b\in F_n^{-1}(1-q)$, the analysis in Section \ref{an auxiliary problem} shows that the candidate bad equilibrium parameterized by $q$ and $b$ vanishes if and only if $D(q)\neq\lambda-b$. Now, take a small $\epsilon>0$, and suppose $\widetilde{F}^*$ assigns positive probability to bonuses larger than $\overline{b}+\epsilon$. Then, we define
\begin{equation*}
    \overline{F}_n(b)=\begin{cases}
        F_n(b) &\text{if }b<\overline{b}+\epsilon; \\
        1 &\text{otherwise}.
    \end{cases}
    \qquad
    \overline{F}^*=\begin{cases}
        \widetilde{F}^*(b) &\text{if }b<\overline{b}+\epsilon; \\
        1 &\text{otherwise}.
    \end{cases}
\end{equation*}
One can see such truncation preserves the weak convergence of $(\overline{F}_n)_{n=1}^{\infty}$ to $\overline{F}^*$, and $\E_{\overline{F}_n}[b]\rightarrow\E_{\overline{F}^*}[b]$. Furthermore, each $\overline{F}_n$ still fully implements full working because: (i) $\supp(\overline{F}_n)\geq\underline{b}$ so a full-working equilibrium exists; (ii) the candidate bad equilibria under $\overline{F}_n$ are the same as under $F_n$ except for those with a threshold bonus $b\geq\overline{b}+\epsilon$, but these cases satisfy $D(q)>\lambda-b$ regardless of $q$ and thus these equilibria still cannot exist. As a result, by Definition \ref{definition-solution concept}, the fact that $\overline{F}^*$ has a strictly lower expectation than $\widetilde{F}^*$ forms a contradiction to $\widetilde{F}^*$ being robustly optimal. In summary, it is without loss to assume $\supp(\widetilde{F}^*)$ and $\supp(F_n)\leq\overline{b}+\epsilon$. Therefore, let $q_H=0$ and take some $b_H>\overline{b}+\epsilon$, which satisfies $b_H\in F_n^{-1}(1-q_H)$ and $b_H\in \widetilde{F}^{*-1}(1-q_H)$. In this case, $b_H+D(q_H)>\lambda$. Now, suppose there is $\widehat{q}\in[0,1)$ and $\widehat{b}\in \widetilde{F}^{*-1}(1-\widehat{q})$ such that $\widehat{b}+D(\widehat{q})<\lambda$. Since the graph $\{(b,q):b\in F_n^{-1}(1-q)\}$ converges to $\{(b,q):b\in \widetilde{F}^{*-1}(1-q)\}$ (with respect to the Hausdorff metric) due to weak convergence, and $D(\cdot)$ is continuous, we can find a large $n$, and some $(b_L,q_L)$ close to $(\widehat{b},\widehat{q})$ such that $b_L\in F_n^{-1}(1-q_L)$ and $b_L+D(q_L)<\lambda$. Therefore, we have the following two nonempty sets
\begin{equation*}
    \begin{aligned}
        A_n&=\{q\in[q_H,q_L]:\exists b\in F_n^{-1}(1-q)\text{ 
s.t. }b+D(q)\geq\lambda\}; \\
        B_n&=\{q\in[q_H,q_L]:\exists b\in F_n^{-1}(1-q)\text{ 
s.t. }b+D(q)\leq\lambda\}.
    \end{aligned}
\end{equation*}
The continuity of $D(\cdot)$ and $F_n^{-1}(\cdot)$ implies that $A_n$ and $B_n$ are both closed sets. Since $A_n\cup B_n=[q_H,q_L]$ by construction, $A_n\cap B_n$ is nonempty. Namely, there are $q'\leq q_L<1$ and $b'\in F_n^{-1}(1-q')$ such that $b'+D(q')=\lambda$. However, this contradicts $F_n$ fully implementing full working as $b'$ and $q'$ parameterize a bad equilibrium. Hence, $\widetilde{F}^*$ satisfies (\ref{equilibrium breaking}) and is feasible in problem (\ref{auxiliary problem}).

Before the second step, we show $F^*(\cdot)$ is strictly increasing on $(\underline{b},\overline{b})$. For every $\widetilde{b}\in(\underline{b},\overline{b})$, let $\widetilde{q}$ be $\overline{q}(\widetilde{b})$. Lemma \ref{lemma-strict increasing} shows that $D(q)\geq D(\widetilde{q})$ for all $q\geq\widetilde{q}$. Take any $\widehat{b}\in(\widetilde{b},\overline{b})$. Hence, every $\widehat{q}$ such that $\widehat{b}+D(\widehat{q})=\lambda$ must have $D(\widehat{q})=\lambda-\widehat{b}<\lambda-\widetilde{b}=D(\widetilde{q})$, so $\widehat{q}$ has to be lower than $\widetilde{q}$. This implies $\overline{q}(\widehat{b})<\overline{q}(\widetilde{b})$. Namely, $\overline{q}(\cdot)$ is strictly decreasing; thus, $F^*=1-\overline{q}$ is strictly increasing.

Also, we characterize $F^*$ by showing it is continuous if and only if $D(\cdot)$ is strictly increasing. When $D(\cdot)$ is strictly increasing, for all $b\in[\underline{b},\overline{b}]$, $\{q:b+D(q)=\lambda\}$ is a singleton that contains $D^{-1}(\lambda-b)$. Hence, $\overline{q}(b)=D^{-1}(\lambda-b)$, continuous in $b$. Moreover, $\overline{q}(\underline{b})=D^{-1}(\lambda-\underline{b})=D^{-1}(D(1))=1$ and $\overline{q}(\overline{b})=D^{-1}(\lambda-\overline{b})=D^{-1}(\min_{q\in[0,1]}D(q))=D^{-1}(D(0))=0$. As a result, $F^*$ is continuous. Conversely, suppose $F^*(\cdot)$ is continuous. This is equivalent to $\overline{q}(\cdot)$ being continuous with $\overline{q}(\underline{b})=1$ and $\overline{q}(\overline{b})=0$. We know $\overline{q}(\cdot)$ is also strictly decreasing, so $\overline{q}^{-1}(\cdot)$ is a well-defined strictly decreasing function. Therefore, $D(\cdot)$ is strictly increasing because for all $q^2,q^1\in[0,1]$ with $q^2>q^1$, we have $D(q^2)=D(\overline{q}(\overline{q}^{-1}(q^2)))=\lambda-\overline{q}^{-1}(q^2)>\lambda-\overline{q}^{-1}(q^1)=D(\overline{q}(\overline{q}^{-1}(q^1)))=D(q^1)$.

\emph{Step 2}. We suppose there is a solution to problem (\ref{auxiliary problem}), denoted as $\widetilde{F}^*$, that is different from $F^*$ given in Theorem \ref{main result}. Note that (\ref{equilibrium keeping}) requires $\widetilde{F}^*(b)=0$ for all $b<\underline{b}$, and that $F^*(b)=1$ for all $b\geq\overline{b}$. As a result, the linear objective of problem (\ref{auxiliary problem}) can obtain at $\widetilde{F}^*$ a value no greater than at $F^*$ only when
\begin{equation*}
    \int_{\underline{b}}^{\overline{b}}\widetilde{F}^*(b)db\geq\int_{\underline{b}}^{\overline{b}}F^*(b)db.
\end{equation*}
Hence, $\widetilde{F}^*$ differs from $F^*$ only when there is some open subset of $[\underline{b},\overline{b})$ in which $\widetilde{F}^*>F^*$. We choose one $\widetilde{b}$ in this set, at which $\widetilde{F}^*$ is continuous. An implication is that $\widehat{b}:=\min F^{*-1}(\widetilde{F}^*(\widetilde{b}))>\widetilde{b}$ since, otherwise, we have $\widetilde{F}^*(\widetilde{b})\leq F^*(\widetilde{b})$. Moreover, $F^*(\widehat{b}^-)\leq\widetilde{F}^*(\widetilde{b})$, so if we take $b'\in\widetilde{F}^{*-1}(F^*(\widehat{b}^-))$, we have $b'\leq\widetilde{b}<\widehat{b}$. Let $q'=1-F^*(\widehat{b}^-)$, and since $F^*(\cdot)$ is strictly increasing on $(\underline{b},\overline{b})$ and $\widehat{b}\in(\underline{b},\overline{b})$, we have $q'<1$. Note that $F^*(\widehat{b}^-)=1-\overline{q}(\widehat{b})$, so $q'=\overline{q}(\widehat{b})$ and thus $\widehat{b}+D(q')=\lambda$. However, a contradiction to (\ref{equilibrium breaking}) occurs as $q'<1$ and $b'\in\widetilde{F}^{*-1}(1-q')$ but $b'+D(q')<\widehat{b}+D(q')=\lambda$. Hence, such an $\widetilde{F}^*$ does not exist. Finally, it is easy to verify that $F^*$ satisfies (\ref{equilibrium keeping}) and (\ref{equilibrium breaking}) by its construction. In this way, we have shown $F^*$ is the unique solution to problem (\ref{auxiliary problem}).

\emph{Step 3}. We show $F^*$ can be approximated by a sequence $(F_n)_{n=1}^{\infty}$ feasible in problem (\ref{minimal cost guarantee}). Since $F^*(\cdot)$ is strictly increasing on $(\underline{b},\overline{b})$, fixing a small $\epsilon>0$, we can construct the weakly approximating sequence so that every $F_n\in\Delta(\mathbb{R_+})$ is continuous on $\mathbb{R_+}$ (that is, atomless), and is strictly below $F^*$ on $(\underline{b},\overline{b}+\epsilon)$ while coinciding with $F^*$ outside this interval (except we fix $F_n(\underline{b})=0$). Such an $F_n$ induces a full-working equilibrium since $F_n(b)=F^*(b)=0$ for all $b<\underline{b}$, namely $\supp(F_n)\geq\underline{b}$. Moreover, for every candidate bad equilibrium parameterized by some $q<1$ and $b\in F_n^{-1}(1-q)$: (i) if $b\in(\underline{b},\overline{b}]$, we have $q=1-F_n(b)>1-F^*(b^-)=\overline{q}(b)$, which along with Lemma \ref{lemma-strict increasing} implies $D(q)>D(\overline{q}(b))$, so $b+D(q)>b+D(\overline{q}(b))=\lambda$; (ii) if $b>\overline{b}$, we always have $b+D(q)>\lambda$. In either case, the bad equilibria cannot exist, so $F_n$ fully implements full working. In addition, weak convergence here implies $\E_{F_n}[b]\rightarrow\E_{F^*}[b]$. As a result, $(F_n)_{n=1}^{\infty}$ approximates $F^*$ in the sense of Definition \ref{definition-solution concept}.

With all three steps above, we have completed the proof. To see this, steps 1 and 3 together show that problems (\ref{minimal cost guarantee}) and (\ref{auxiliary problem}) cap each other's value, so their values coincide. Next, steps 2 and 3 together show $F^*$ achieves the optimal value of problem (\ref{auxiliary problem}) and thus that of problem (\ref{minimal cost guarantee}), while being approximated by a sequence feasible in problem (\ref{minimal cost guarantee}), so it is a robustly optimal incentive policy. Finally, suppose there is another robustly optimal incentive policy $\widetilde{F}^*$, then step 1 implies it is feasible in problem (\ref{auxiliary problem}) and thus optimal there. However, step 2 says that problem (\ref{auxiliary problem}) has a unique solution, so we reach a contradiction. In summary, $F^*$ is the unique robustly optimal incentive policy (Theorem \ref{main result}), and problems (\ref{minimal cost guarantee}) and (\ref{auxiliary problem}) coincide with each other in their solutions (Proposition \ref{proposition-auxiliary problem}).

\subsection{Proof of Proposition \ref{proposition-occupational mobility} and Proposition \ref{proposition-assortativeness skill premium}}

We begin by showing the first part of Proposition \ref{proposition-assortativeness skill premium}. Note the following: (i) $\E_0[p_k^1]=\E_0[p_k^2]$ controls for the first term of (\ref{career value-linear environment}), $T(1;\E_0[p_k^1])=T(1;\E_0[p_k^2])$; (ii) the fact that $F^{*2}$ contains contract dispersion implies $\Cov(u^2,p^2)>0$. If $\Cov(u^1,p^1)\leq0$, $F^{*1}$ is degenerate and the result holds trivially; henceforth, we consider $\Cov(u^1,p^1)>0$. Since $\Cov(u^2,p^2)>\Cov(u^1,p^1)>0$, there is $\eta\in[0,1)$ such that the career value functions have $D^1=\eta D^2$ (in Proposition \ref{proposition-occupational mobility}, the condition $v^1=\delta v^2$ also implies this with $\eta$ equal to $\delta$, so the following proof applies there as well). Then, we have $\overline{b}^1-\underline{b}^1=D^1(1)-\min_qD^1(q)=\eta D^2(1)-\min_q\eta D^2(q)=\eta(\overline{b}^2-\underline{b}^2)<\overline{b}^2-\underline{b}^2$. Moreover, for all $\Delta b\in(0,\overline{b}^2-\underline{b}^2)$, we have $\eta\Delta b\in(0,\overline{b}^1-\underline{b}^1)$, and
\begin{equation}\label{q=q}
    \begin{aligned}
        \overline{q}^1(\underline{b}^1+\eta\Delta b)&=\max\{q:\underline{b}^1+\eta\Delta b+D^1(q)=\lambda\}=\max\{q:\eta\Delta b+D^1(q)=D^1(1)\} \\
        &=\max\{q:\eta\Delta b+\eta D^2(q)=\eta D^2(1)\}=\max\{q:\Delta b+D^2(q)=D^2(1)\} \\
        &=\max\{q:\underline{b}^2+\Delta b+D^2(q)=\lambda\}=\overline{q}(\underline{b}^2+\Delta b).
    \end{aligned}
\end{equation}
Above, the second and fifth equalities apply $\underline{b}^i=\lambda-D^i(1)$ for both $i=1,2$. Equation (\ref{q=q}) implies that, if we let $X^i$ denote a random variable distributed as $F^{*i}$ for both $i=1,2$, we have $X^1-\underline{b}^1$ equals $\eta(X^2-\underline{b}^2)$ in distribution, denoted $X^1-\underline{b}^1\stackrel{d}{=}\eta(X^2-\underline{b}^2)$. We now use this to verify the convex-order condition in Definition \ref{definition-more dispersed}. Subtracting means on both sides yields
\begin{equation*}
    X^1-\E_{F^{*1}}[b]\;=\;\eta\bigl(X^2-\E_{F^{*2}}[b]\bigr).
\end{equation*}
Set $Y:=X^2-\E_{F^{*2}}[b]$, so $\E[Y]=0$. For any convex $\phi:\mathbb{R}\to\mathbb{R}$ and any $y\in\mathbb{R}$, convexity gives
\begin{equation*}
    \phi(\eta y)\;=\;\phi\bigl(\eta y+(1-\eta)\cdot 0\bigr)\;\le\;\eta\,\phi(y)+(1-\eta)\,\phi(0),
\end{equation*}
and Jensen's inequality gives $\phi(0)=\phi(\E[Y])\le\E[\phi(Y)]$. Taking expectations and combining the two,
\begin{equation*}
    \E[\phi(\eta Y)]\;\le\;\eta\,\E[\phi(Y)]+(1-\eta)\,\phi(0)\;\le\;\E[\phi(Y)].
\end{equation*}
Translating back,
\begin{equation*}
    \E_{F^{*1}}\bigl[\phi(b-\E_{F^{*1}}[b])\bigr]\;\le\;\E_{F^{*2}}\bigl[\phi(b-\E_{F^{*2}}[b])\bigr].
\end{equation*}
Since this holds for every convex $\phi$, Definition \ref{definition-more dispersed} yields that $F^{*2}$ is more dispersed than $F^{*1}$.

Next, we show the second part of Proposition \ref{proposition-assortativeness skill premium}. We assume $p=p^1=p^2$, $\E_0[u_k^1]=\E_0[u_k^2]$, and that there is $k^*\in K$ such that $u_k^2>[<]u_k^1$ for all $p_k>[<]p_{k^*}$. Then, the covariances differ by the following
\begin{equation*}
    \begin{aligned}
        \Cov(u^2,p)&-\Cov(u^1,p)=\E_0[u_k^2p_k]-\E_0[u_k^1p_k]=\sum_{k\in K}\mu^0_kp_k(u_k^2-u_k^1) \\
        &=\sum_{k\in K:p_k>p_{k^*}}\mu^0_kp_k(u_k^2-u_k^1)+\sum_{k\in K:p_k<p_{k^*}}\mu^0_kp_k(u_k^2-u_k^1)+\sum_{k\in K:p_k=p_{k^*}}\mu^0_kp_{k^*}(u_k^2-u_k^1) \\
        &>\sum_{k\in K:p_k>p_{k^*}}\mu^0_kp_{k^*}(u_k^2-u_k^1)+\sum_{k\in K:p_k<p_{k^*}}\mu^0_kp_{k^*}(u_k^2-u_k^1)+\sum_{k\in K:p_k=p_{k^*}}\mu^0_kp_{k^*}(u_k^2-u_k^1) \\
        &=\sum_{k\in K}\mu^0_kp_{k^*}(u_k^2-u_k^1)=\left(\E_0[u_k^2]-\E_0[u_k^1]\right)p_{k^*}=0.
    \end{aligned}
\end{equation*}
Above, we have a strict inequality because we have assumed $\Cov(u^2,p)>0$, which means $p_k$ cannot be ability-independent, so there is some $k\neq k^*$ such that $p_k\neq p_{k^*}$, and we have either $\sum_{k\in K:p_k>p_{k^*}}\mu^0_kp_k(u_k^2-u_k^1)>\sum_{k\in K:p_k>p_{k^*}}\mu^0_kp_{k^*}(u_k^2-u_k^1)$ or $\sum_{k\in K:p_k<p_{k^*}}\mu^0_kp_k(u_k^2-u_k^1)>\sum_{k\in K:p_k<p_{k^*}}\mu^0_kp_{k^*}(u_k^2-u_k^1)$. The proposition then follows.

\newpage

\section{Supplemental Appendix}

\subsection{Classic Monitoring System}\label{classic monitoring system}

This section studies a one-period version of the classic contracting-with-career-concerns setup considered by, for example, \cite{gimu1992}, and shows that with the additive monitoring system they assume, every contract induces a unique equilibrium. In addition, by using this one-period result, even in a multiple-period model, one obtains uniqueness by working backward from the last period to the first period.
The timing coincides with our main model. However, the worker now chooses a continuous action $a\geq0$, and the project's outcome can be any real number $y\in\mathbb{R}$. The worker's ability level $k\in\mathbb{R}$ is normally distributed with mean $m_0$ and variance $\sigma_0^2$. The monitoring system is additive and takes the following linear form
\begin{equation*}
    y=a+k+\epsilon,
\end{equation*}
where the noise term $\epsilon$ is independent of ability, and normally distributed with mean 0 and variance $\sigma_{\epsilon}^2$. We consider the equilibria (PBE) induced by any contract $b(y)$ that specifies the outcome-contingent transfer to the worker. Moreover, we also do not impose any restriction on the post-employment value $v(\cdot)$, which can be viewed as a continuation value in a given multiple-period equilibrium. Since we allow $b(\cdot)$ and $v(\cdot)$ to be arbitrary, we are considering a setting that is potentially more general than \cite{gimu1992}, who consider perfect firm competition and linear contracts. Because we discuss the strategic uncertainty facing a partially implementing contract, it suffices to consider a deterministic one. We denote the worker's effort cost by $c(a)$.

We begin by characterizing the worker's career concerns when the market expects him to choose some action $\hat{a}$. Given this expectation, well-known filtering formulas (e.g., \cite{degr2005}) yield that conditional on observing outcome $y$, the market's posterior belief about ability $k$ is a normal distribution with
\begin{equation*}
    \text{mean }\frac{\sigma_{\epsilon}^2m_0+\sigma_0^2(y-\hat{a})}{\sigma_{\epsilon}^2+\sigma_0^2}\text{ and variance }\frac{\sigma_0^2\sigma_{\epsilon}^2}{\sigma_0^2+\sigma_{\epsilon}^2}.
\end{equation*}
Thus, the post-employment value, which is a function of the posterior, can be rewritten as a reduced-form function of $y-\hat{a}$ alone, denoted simply as $v(y-\hat{a})$. Note that when the market believes the worker chooses $\hat{a}$ while he actually chooses $a$, $y-\hat{a}$ is normally distributed with mean $m_0+a-\hat{a}$ and variance $\sigma^2:=\sigma_0^2+\sigma_{\epsilon}^2$. Hence, the worker's future expected value in the market is
\begin{equation*}
    U(a|\hat{a})=\int_{\mathbb{R}}\frac{v(s)}{\sigma}\phi\left(\frac{s-(m_0+a-\hat{a})}{\sigma}\right)ds\text{, where $\phi$ is the standard Gaussian p.d.f.}
\end{equation*}
Notice that $U(a|\hat{a})=U(a-\hat{a}|0)$. We let $U_0(a):=U(a|0)$ and thus $U(a|\hat{a})=U_0(a-\hat{a})$.

Moreover, the worker's payoff from the employer when choosing $a$ is
\begin{equation*}
    W(a)=\int_{\mathbb{R}}\frac{b(s)}{\sigma}\phi\left(\frac{s-(m_0+a)}{\sigma}\right)ds.
\end{equation*}
Consequently, an equilibrium where the worker chooses $\hat{a}$ exists if and only if $\hat{a}$ is the worker's optimal choice given the market belief that it is also his actual decision
\begin{equation}\label{worker problem}
    \hat{a}\in\argmax_{a\geq0}W(a)-c(a)+U_0(a-\hat{a}).
\end{equation}
Note that $\hat{a}$ enters payoffs only through $a-\hat{a}$.
More specifically, we follow \cite{gimu1992} and consider any linear contract $b(\cdot)$, which implies linear $W(\cdot)$. In addition, we consider strictly convex effort cost $c(\cdot)$. The equilibrium is then uniquely pinned down by the first-order condition for (\ref{worker problem}): $W'(\hat{a})-c'(\hat{a})+U_0'(0)=0$. Career concerns enter this condition only through a constant $U_0'(0)$ that does not vary with the market expectation $\hat{a}$. When one compares this condition with (\ref{equilibrium concept}), roughly speaking, $U_0'(0)$ is the ``counterpart'' of the career value there. Hence, the additive setup fails to incorporate effort's influence on the outcome-generating process.

\subsection{Informed Worker: Omitted Analysis and Generalizations}\label{private types: partial working and multiple types}

\paragraph{Omitted Analysis of Section \ref{informed worker}.} For the binary-type full-working case, we derive the equilibrium-keeping and equilibrium-breaking constraints to write an auxiliary problem. For notational convenience, we associate each incentive policy $F$ with its \emph{tail incentive policy} $R(b):=1-F(b^-)$. $R(b^*)=\Pr_F(b\geq b^*)$ is the total working probability if the worker works with probability one at all bonuses no less than threshold $b^*$. Let $\Delta_T(\mathbb{R}_+)$ be the space of tail policies. The first constraint below, equilibrium-keeping, asserts that the employer should assign all probability to bonuses no less than the partial-implementation bonus $\underaccent{\sim}{b}$
\begin{equation}\label{equilibrium keeping-private type}
    R(\underaccent{\sim}{b})=1. \tag{EK$'$}
\end{equation}
By definition, (\ref{equilibrium keeping-private type}) is equivalent to $F(\underaccent{\sim}{b}^-)=0$. The second constraint, equilibrium-breaking, applies the following property of each candidate equilibrium: both types' threshold bonuses differ by a constant, that is, the high type is indifferent between working and shirking at bonus $b_H=\lambda_H-D$ and the low type at $b_L=\lambda_L-D$ where $D$ is the equilibrium career value, which implies $b_L-b_H=\lambda_L-\lambda_H$. Let $\lambda_0:=\lambda_L-\lambda_H$ denote the \emph{cost gap}. One can see the worker's strategy profile is pinned down (except for the decision at thresholds) by a scalar. The equilibrium-breaking constraint then comprises a set of conditions for breaking a one-dimensional continuum of bad equilibria
\begin{equation}\label{equilibrium breaking-private type}
    \text{ For all $b\geq \underaccent{\sim}{b}-\lambda_0$, }b+D(R(b+\lambda_0),R(b))\geq\lambda_H. \tag{EB$'$}
\end{equation}
To understand this condition, consider a candidate equilibrium with high-type threshold $b_H=b$, which implies the low type's threshold bonus $b_L=b+\lambda_0$. Thus, (\ref{equilibrium breaking-private type}) demands that in this equilibrium indexed by $b$, the market's beliefs formed by anticipating the high type's total working probability $R(b)$ and the low type's total working probability $R(b+\lambda_0)$ must generate a career value $D(R(b+\lambda_0),R(b))$ that is too high to rationalize the anticipated worker behavior. That is, given this induced career value, both types would like to work more than expected. Notice that in a desired full-working equilibrium, the high type's threshold bonus is $\lambda_H-D(1,1)=\underaccent{\sim}{b}-\lambda_0$, so the bad equilibria are indexed by $b>\underaccent{\sim}{b}-\lambda_0$.

We show (\ref{equilibrium keeping-private type}) and (\ref{equilibrium breaking-private type}) are also sufficient for finding the robustly optimal incentive policy.

\begin{proposition}\label{proposition-auxiliary problem-private type}
    $F^{\dagger}$ is robustly optimal if and only if its associated tail incentive policy $R^{\dagger}$ solves the following problem
    \begin{equation}\label{auxiliary problem-private type}
        \min_{R\in\Delta_T(\mathbb{R_+})}\int_{\underaccent{\sim}{b}}^{\infty}R(b)db\qquad\text{s.t. (\ref{equilibrium keeping-private type}) and (\ref{equilibrium breaking-private type})}.
    \end{equation}
\end{proposition}

The key to solving (\ref{auxiliary problem-private type}) again relies on a greediness result: the employer finds it optimal to assign as little probability to high bonuses as is needed. Formally, (\ref{equilibrium breaking-private type}) binds. We give a name, $b$-EB', to the equilibrium-breaking constraint indexed by $b$. To understand why (\ref{equilibrium breaking-private type}) optimally binds, take a $b>\underaccent{\sim}{b}$ (so that (\ref{equilibrium keeping-private type}) is not restrictive at $b$) with $R(b)>0$, and suppose $b$-EB' is slack. However, the employer can be strictly better off by lowering $R$ around $b$ because: (i) this reduces the cost $\int R(b)db$; (ii) the resulting tail incentive policy still satisfies (\ref{equilibrium breaking-private type}) and thus remains feasible in problem (\ref{auxiliary problem-private type}). To see (ii), we notice that the value $R(b)$ enters two EB' constraints, $b$-EB' and $(b-\lambda_0)$-EB', as follows
\begin{equation*}
    \begin{aligned}
        b+D(R(b+\lambda_0),R(b))&\geq\lambda_H; \\
        (b-\lambda_0)+D(R(b),R(b-\lambda_0))&\geq\lambda_H.
    \end{aligned}
\end{equation*}
Because $b$-EB' is slack, a small perturbation of $R(b)$ will not violate this constraint. On the other hand, $R(b)$ enters $(b-\lambda_0)$-EB' as the total working probability of the low type, so lowering $R(b)$ will increase $D(R(b),R(b-\lambda_0))$, and $(b-\lambda_0)$-EB' is not violated. We hence show the following result

\begin{theorem*}[\ref{main result-private type}]
    To induce full working from a binary-type informed worker, the unique robustly optimal incentive policy $F^{\dagger}$ is given by
    \begin{equation*}
        \supp(F^{\dagger})=[\underaccent{\sim}{b},\accentset{\sim}{b}]\text{, and }b\text{-EB' binds for all }b\in(\underaccent{\sim}{b},\accentset{\sim}{b}].
    \end{equation*}
    Moreover, $F^{\dagger}$ is continuous on $(\underaccent{\sim}{b},\accentset{\sim}{b}]$ and has a mass point at $\underaccent{\sim}{b}$.
\end{theorem*}

\paragraph{General Setting.} Next, we define the robust solution concept for fully implementing partial working when the worker has multiple types, and present the general results. To begin with, we label the ability types as integers $ K=\{1,2,...,K\}$ and fix a profile $Q\in[0,1]^K$ such that the $k$th coordinate, $Q_k$, refers to each type $k$'s total working probability. We assume that the effective cost $\lambda_k$ strictly decreases with the type label $k$, so higher types are more skilled in the sense that their efforts produce success more effectively. Given the equilibrium concept specified in Section \ref{informed worker}, we collect all incentive policies that uniquely induce $Q$ in equilibrium in $\mathcal{F}^{FI}(Q)$. Hence, for every $g\in\mathcal{E}(F)$ with $F\in\mathcal{F}^{FI}(Q)$, $\E_F[\sigma_k(b)]=Q_k$ for all $k$. For now, we take as granted that $Q$ is implementable, that is $\mathcal{F}^{FI}(Q)\neq\emptyset$, and return to implementability later. The following then defines the employer's objective conditional on her goal of inducing $Q$ robustly.

\begin{definition}\label{definition-solution concept-Q}
    \emph{The employer's }minimal cost guarantee\emph{ for fully implementing $Q$ is
    \begin{equation}\label{minimal cost guarantee-Q}
        W^{\dagger}_Q=\inf_{F\in\mathcal{F}^{FI}(Q)}\sup_{g\in\mathcal{E}(F)}\E_F\left[b\sum_{k\in K}\mu^0_k(p_{0k}+p_k\sigma_k(b))\right].
    \end{equation}
    An incentive policy $F^{\dagger}_Q$ is }robustly optimal\emph{ if there is a sequence of incentive policies $(F_n)_{n=1}^{\infty}$ such that:
    \begin{enumerate}[nolistsep]
        \item The sequence $(F_n)_{n=1}^{\infty}$ is contained in $\mathcal{F}^{FI}(Q)$ and weakly converges to $F^{\dagger}_Q$;
        \item The cost guarantee, $\sup_{g\in\mathcal{E}(F_n)}\E_{F_n}\left[b\sum_{k\in K}\mu^0_k(p_{0k}+p_k\sigma_k(b))\right]$, converges to $W^{\dagger}_Q$.
    \end{enumerate}}
\end{definition}

Note that the above program entails an assumption that the employer always expects the worst-case equilibrium in $\mathcal{E}(F)$. However, this selection rule is not essential because each type uses a threshold strategy in equilibrium, so $Q$ and $F$ suffice to pin down the expected cost.

Below we present the extension results, whose proofs can be found in later sections.

\paragraph{Binary Types with Partial Working.} First, we investigate the full implementation of partial working of a binary-type worker. The assumption needed is the same as the one made in Section \ref{informed worker}, namely, $v(\cdot)$ is continuous and strictly increasing in the high type's probability. We also borrow the notations there: $ K=\{H,L\}$, career value $D:[0,1]^2\rightarrow\mathbb{R}$, and tail incentive policy $R$. 

The analysis again starts with the following critical bonuses
\begin{equation*}
    \underline{b}_L=\lambda_L-D(Q_L,Q_H);\quad\quad\quad\underline{b}_H=\lambda_H-D(Q_L,Q_H);\quad\quad\quad\accentset{\sim}{b}=\lambda_H-D(0,0).
\end{equation*}
Note that $\underline{b}_k$ denotes type $k$'s on-path threshold bonus while $\accentset{\sim}{b}$ refers to the bonus that guarantees the high type's work. As in the full-working case, the employer's optimal strategy here is greedy, iteratively breaking bad outcomes by inducing small deviations. To formalize greediness, we first define a tail incentive policy's support as its associated incentive policy's support, namely $\supp(R):=\supp(F)$. We then pin down the greedy behavior of the employer, which yields binding (\ref{equilibrium breaking-private type}).

\begin{definition}\label{definition-greedy policy}
    \emph{The} greedy incentive policy\emph{ $R^G\in\Delta_T(\mathbb{R}_+)$ is defined by the following:
    \begin{enumerate}[nolistsep]
        \item $b$-EB' holds if $R^G(b)<1$;
        \item $b$-EB' binds if and only if $b\in\supp(R^G)$.
    \end{enumerate}}
\end{definition}

It turns out that $R^G$ can be uniquely pinned down. We first show that the upper bound of its support must be $\accentset{\sim}{b}$. Next, proceeding from this bonus to lower ones, we use the binding EBs to back out the entire function given that $R^G=0$ at bonuses higher than $\accentset{\sim}{b}$. The construction stops when the function hits the boundary $R^G=1$. We summarize a few properties of $R^G$ below

\begin{lemma}\label{lemma-greedy solution}
    $R^G$ is unique, continuous, and fully supported on $[b_l,\accentset{\sim}{b}]$ with some $b_l<\underaccent{\sim}{b}$.
\end{lemma}

Next, we construct the $Q$-equilibrium-breaking policy $R^G_Q$ and the $Q$-equilibrium-keeping policy $R^K_Q$, each of which translates the corresponding constraints (EB/EK) into the employer's behavior.
\begin{equation*}
    \begin{aligned}
        R^G_Q(b)&=\min\{R^G(b),Q_H\}. \\
        R^K_Q(b)&=\begin{cases}
        1&\text{ if }b=0; \\
        Q_H&\text{ if }b\in(0,\underline{b}_H]; \\
        Q_L&\text{ if }b\in(\underline{b}_H,\underline{b}_L]; \\
        0&\text{ if }b>\underline{b}_L.
    \end{cases}
    \end{aligned}
\end{equation*}
Notice that $R_Q^G$ basically represents the greedy strategy $R^G$ while being capped by $Q_H$ because $Q_H$ is the maximal working probability the employer induces for the high type. Moreover, $R_Q^K$ captures the need to maintain the existence of the desired equilibrium where the career value is $D(Q_L,Q_H)$ and each type $k$'s threshold bonus is $\underline{b}_k$. Given these, $R_Q^K$ ensures that type $k$'s working probability is exactly $Q_k$.

Now, we are ready to formally present the extension result

\begin{proposition}\label{main result-Q}
    To induce $Q$ from a binary-type informed worker, the unique robustly optimal incentive policy $F^{\dagger}_Q$ is associated with the following tail incentive policy, for all $b\geq0$
    \begin{equation}\label{ROWP-Q}
        R_Q^{\dagger}(b)=\max\{R_Q^K(b),R_Q^G(b)\}.
    \end{equation}
    Moreover, $F^{\dagger}_Q$ takes one of the three forms below. If $Q_H>Q_L>0$, $F_Q^{\dagger}$ contains two mass points at $\underline{b}_H$ and $\underline{b}_L$, and there is $\widehat{b}\in[\underline{b}_H,\underline{b}_L)$ such that $F_Q^{\dagger}$ is fully supported and continuous on $(\underline{b}_H,\widehat{b}]$ and $(\underline{b}_L,\accentset{\sim}{b}]$. If $Q_H=Q_L>0$, $F^{\dagger}_Q$ contains a mass point at $\underline{b}_L$ and is fully supported and continuous on $(\underline{b}_L,\accentset{\sim}{b}]$. If $Q_H\geq Q_L=0$, $F^{\dagger}_Q$ is fully supported and continuous on $[\underline{b}_H,\accentset{\sim}{b}]$.
\end{proposition}

\begin{figure}[h]
    \centering
    \includegraphics[width=1\linewidth]{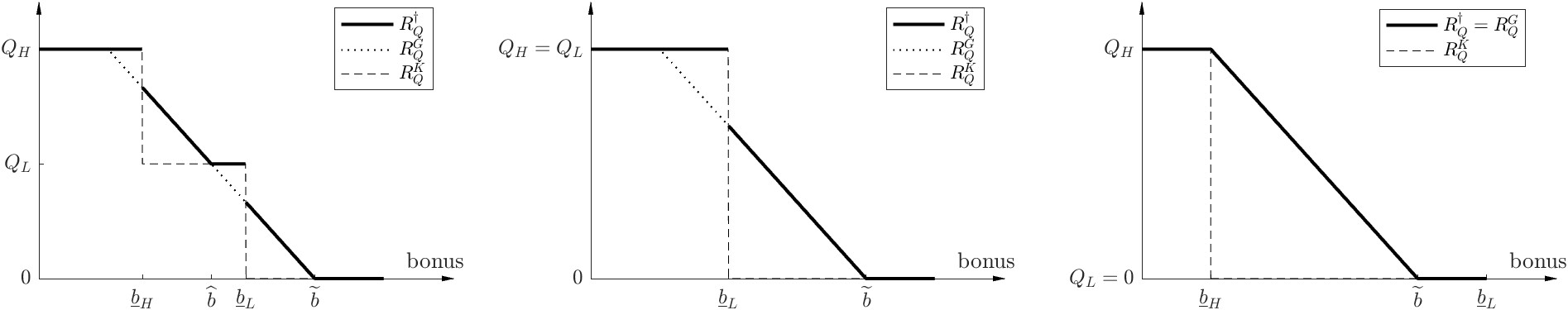}
    \caption{Partial working of a binary-type worker.}
    \label{fig-partial working of binary type}
\end{figure}

Proposition \ref{main result-Q} extends the full-working result, Proposition \ref{main result-private type}, and highlights the optimality of greediness as (\ref{ROWP-Q}) demonstrates that everywhere in the bonus range, either EK or EB must bind. Figure \ref{fig-partial working of binary type} depicts how the optimal contract structure varies with $Q$. The interesting case is where $Q_H>Q_L>0$. Recall that in the full-working case, the optimal design takes on a dispersion-above-mass contract structure, that is, the employer announces a minimum bonus $\underaccent{\sim}{b}$ while privately offering dispersed bonus raises to some workers. When $Q_H>Q_L>0$, she instead uses two dispersion-above-mass structures. In particular, she randomly divides workers into two groups and announces a minimum bonus $\underline{b}_H$ [$\underline{b}_L$] in the first [second] group. Moreover, she privately offers dispersed bonus raises to some workers in each group.

Finally, we provide the sufficient and necessary criterion for implementability of $Q$

\begin{proposition}\label{proposition-implementability}
    For $Q_H\geq Q_L$, $Q$ is not implementable if and only if $Q_L\leq R^G(\underline{b}_L)$ and $\underline{b}_L\in\supp(R^G)$.
\end{proposition}

In short, Proposition \ref{proposition-implementability} shows that for each fixed $Q_H$, there is a lower bound $\underline{Q}_L$ such that $Q$ is implementable if and only if $Q_L\in(\underline{Q}_L,Q_H]$. On the one hand, we must have $Q_L\leq Q_H$ because the high type is willing to work at all bonuses where the low type works; on the other hand, $Q_L$ cannot be lower than $\underline{Q}_L$ because the need to break bad equilibria forces the employer to assign probabilities to high bonuses (as in the greedy policy $R^G$), which guarantees a minimal working probability of the low type, $\underline{Q}_L$.

\paragraph{Multiple Types with Full Working.} Next, we investigate the full implementation of full working of a finite-type worker. Apart from the continuity of $D(\cdot)$, we assume that $D(\cdot)$ is quasi-concave in $q_k$ for each $k$. One can see that this extends the binary-type assumption that $D(\cdot)$ increases [decreases] strictly in $q_H$ [$q_L$], as a result of $v(\cdot)$ strictly increasing in high-type probability. We are then able to derive the sufficient and necessary condition for strategic uncertainty and contract dispersion

\begin{proposition}\label{proposition-criterion for contract dispersion-general}
    To induce full working, contract dispersion is robustly optimal if and only if
    \begin{equation}\label{criterion for contract dispersion-general}
        \underaccent{\sim}{b}:=\lambda_1-D(1,...,1)<\accentset{\sim}{b}:=\max\left\{\max_{k\in K}\lambda_k-D(\underbrace{0,...,0}_{\text{$k$ zeros}},1,...,1),\underaccent{\sim}{b}\right\}.
    \end{equation}
    Moreover, the deterministic robust bonus is $\accentset{\sim}{b}$ and the partial-implementation policy is degenerate at $\underaccent{\sim}{b}$.
\end{proposition}

Proposition \ref{proposition-criterion for contract dispersion-general} preserves the paper's main message, which connects incentive provision with the employer's goal of ruling out low-effort beliefs. However, fully characterizing the robustly optimal incentive policy is difficult when the worker has multiple types. The reason is that greediness may no longer be optimal. Specifically, the EB constraints in this case are
\begin{equation}\label{equilibrium breaking-multiple type}
    \text{ For all $b\geq\underaccent{\sim}{b}-(\lambda_1-\lambda_K)$, }b+D((R(b+\lambda_k-\lambda_K))_{k=1}^K)\geq\lambda_K. \tag{EB$''$}
\end{equation}
Two facts complicate solving the employer's problem: first, how career value depends on different types' working probabilities can be complex; second, each value $R(b)$ enters at most $K$ constraints, so (\ref{equilibrium breaking-multiple type}) contains interdependent conditions.

\subsection{Proof of Proposition \ref{proposition-auxiliary problem-private type}}

We show a generalized version of Proposition \ref{proposition-auxiliary problem-private type} for the case with multiple types $ K=\{1,2,...,K\}$ and an arbitrary implementable working probability profile $Q=(Q_k)_{k\in K}$. To be consistent with the notations in Appendix \ref{private types: partial working and multiple types}, let $\underline{b}_k:=\lambda_k-D(Q)$, $\underline{b}_0:=\infty$, $\underline{b}_0Q_0:=0$, and $\underline{b}_{K+1}:=0$. For all $R\in\Delta_T(\mathbb{R}_+)$, define
\begin{equation*}
    \begin{aligned}
        V_Q(R)&:=\sum_{k=1}^{K+1}r_k\int_{\underline{b}_k}^{\underline{b}_{k-1}}R(s)ds\text{, where } \\
        r_k&:=\sum_{j=1}^{k-1}\mu^0_jp_{0j}+\sum_{j=k}^K\mu^0_j(p_{0j}+p_j)\text{ for each }k.
    \end{aligned}
\end{equation*}
Here, $\sum_{j=1}^0=0$ and $\sum_{j=K+1}^K=0$. In addition, we introduce a constant $C_Q$ associated with each $Q$
\[C_Q:=\sum_{k=1}^{K+1}r_k(\underline{b}_{k}Q_{k}-\underline{b}_{k-1}Q_{k-1}).\]
We then establish a lower bound of the minimal cost guarantee
\begin{proposition}\label{proposition-auxiliary problem-general}
    For an implementable $Q$, the minimal cost guarantee $W^{\dagger}_Q$ satisfies
    \begin{equation}\label{auxiliary problem-general}
        \begin{aligned}
            W^{\dagger}_Q\geq C_Q+\min_{R}&\text{ }V_Q(R) \\
            \text{s.t.}\quad\emph{[EK]}&\text{ For all $k\in K$, }R(\underline{b}_k)\geq Q_k\text{ and }R(\underline{b}_k^+)\leq Q_k; \\
            \emph{[EB-r]}&\text{ For all $b>\underline{b}_K$, }b+D((R(b+\lambda_k-\lambda_K))_{k=1}^K)\geq\lambda_K; \\
            \emph{[EB-l]}&\text{ For all $b<\underline{b}_K$, }b+D((R(b+\lambda_k-\lambda_K))_{k=1}^K)\leq\lambda_K; \\
            \emph{[FE]}&\text{ $R$ is nonincreasing and left-continuous, and }R\in[0,1].
        \end{aligned}
    \end{equation}
    Moreover, every robustly optimal tail incentive policy $R^{\dagger}_Q$ is feasible in this problem and satisfies $W^{\dagger}_Q=C_Q+V_Q(R^{\dagger}_Q)$. In turn, a solution to this problem is robustly optimal if it can be approximated by a sequence of incentive policies in the sense of Definition \ref{definition-solution concept-Q}.
\end{proposition}

Henceforth, we refer to the four constraints in Problem (\ref{auxiliary problem-general}) as [EK], [EB-r], [EB-l], and [FE]. Problem (\ref{auxiliary problem-general}) provides a lower bound for the minimal cost guarantee. While it is generally not equivalent to the original problem (\ref{minimal cost guarantee-Q}), Proposition \ref{proposition-auxiliary problem-general} establishes the criterion for building such an equivalence. That is, we first obtain a solution to the relaxed problem and next verify it is also ``feasible'' in the original problem. At the end of the proof, we use this result to show Proposition \ref{proposition-auxiliary problem-private type}.

To show Proposition \ref{proposition-auxiliary problem-general}, we start by proving a useful lemma

\begin{lemma}\label{lemma-pure implementation}
    For every robustly optimal incentive policy $R^{\dagger}_Q$, there exists a sequence $(R_n)_{n=1}^{\infty}$ converging to $R^{\dagger}_Q$ in the sense of Definition \ref{definition-solution concept-Q} such that every $R_n$ fully implements $Q$ in the unique pure-strategy equilibrium where each type works with probability one at his threshold bonus $\underline{b}_k$.
\end{lemma}

To show Lemma \ref{lemma-pure implementation} is true, we take any robustly optimal $R^{\dagger}_Q$ and its approximating sequence $(R_n)_{n=1}^{\infty}$. Our goal is to construct another sequence $(\widetilde{R}_n)_{n=1}^{\infty}$ that also approximates $R^{\dagger}_Q$, and each $\widetilde{R}_n$ fully implements $Q$ in a unique pure-strategy equilibrium. In fact, whenever some $R$ fully implements $Q$, it must induce a unique equilibrium because on the equilibrium path, each type $k$'s threshold bonus is fixed as $\underline{b}_k$ and the probability of him working at $\underline{b}_k$ is pinned down as $1-\frac{R(\underline{b}_k)-Q_k}{R(\underline{b}_k)-R(\underline{b}_k^+)}$. Notice that some type $k$ may randomize only because $R(\underline{b}_k)>Q_k\geq R(\underline{b}_k^+)$. Therefore, we simply need to make the potentially mixed equilibrium induced by each $R_n$ pure, by perturbing the values of $R_n$ around $\underline{b}_k$ for some $k$ such that $R_n(\underline{b}_k)>Q_k$. To do so, we consider a sequence of sufficiently small positive numbers $(\epsilon_n)_{n=1}^{\infty}$ that converges to 0. Then, we construct each $\widetilde{R}_n$ in the following way
\begin{equation}\label{purification}
    \widetilde{R}_n(b)=\begin{cases}
        Q_k-\frac{R_n(\underline{b}_k-\epsilon_n)-Q_k}{\epsilon_n}(b-\underline{b}_k) & \text{if }b\in[\underline{b}_k-\epsilon_n,\underline{b}_k]\text{, for all $k$;} \\
        R_n(b) & \text{otherwise.}
    \end{cases}
\end{equation}
In other words, instead of having a jump at $\underline{b}_k$, we linearize the function on the small interval $[\underline{b}_k-\epsilon_n,\underline{b}_k]$ to maintain $\widetilde{R}_n(\underline{b}_k)=Q_k$. As a result, $\widetilde{R}_n$ induces $Q$ in one pure-strategy equilibrium where each type works with probability one at his threshold bonus $\underline{b}_k$. In addition, whenever $R_n(b)$ converges to $R^{\dagger}_Q(b)$ pointwise, which covers all continuous points $b$ of $R^{\dagger}_Q$, $\widetilde{R}_n(b)$ converges pointwise to the same value as $R_n(b)$ does. Thus, $\widetilde{R}_n$ also weakly converges to $R^{\dagger}_Q$.

Now, we show each $\widetilde{R}_n$ fully implements $Q$. Consider any undesirable candidate equilibrium where each type $k$'s threshold bonus is $b'_k$. Let $D^0$ be the career value generated by the worker's strategy in this candidate equilibrium. If $\widetilde{R}_n(b'_k)=R_n(b'_k)$ for all $k$, this candidate equilibrium cannot exist because it is broken under $R_n$. Thus, we consider a candidate equilibrium where some type $k$'s threshold bonus lies in some open linearization interval $(\underline{b}_k-\epsilon_n,\underline{b}_k)$. We collect all such types in $\widetilde{ K}$. In fact, by choosing a sufficiently small $\epsilon_n$, we can make sure that, in every such equilibrium, each type's threshold bonus is a continuous point of $\widetilde{R}_n$. This is because we have finite types, so these threshold bonuses can potentially come from finitely many intervals of the form, $(b'-\epsilon_n,b')$ where $b'=\underline{b}_{k'}+\lambda_{k'}-\lambda_{k''}\text{ for some }k',k''$. It is possible to find an $\epsilon_n$ such that $\widetilde{R}_n$ is continuous on all these intervals. Moreover, suppose two types' threshold bonuses, say $b'_{k'}$ and $b'_{k''}$, are in two linearization intervals $(b^1-\epsilon_n,b^1)$ and $(b^2-\epsilon_n,b^2)$, respectively. By choosing a small $\epsilon_n$, we can make sure that this happens only if $b^1-b^2=\lambda_{k'}-\lambda_{k''}$. Hence, when some threshold bonus $b'_k$ lies in some linearization interval $(\underline{b}_{k'}-\epsilon_n,\underline{b}_{k'})$, denote by $e:=\underline{b}_{k'}-b'_k$ its downward deviation from $\underline{b}_{k'}$, and let $\gamma_{k'}:=\frac{\widetilde{R}_n(b'_k)-R_n(\underline{b}_{k'}^+)}{R_n(\underline{b}_{k'})-R_n(\underline{b}_{k'}^+)}$. Now, consider an auxiliary equilibrium under $R_n$ where each type $k$'s threshold bonus is $b'_k+e$, and if $k\in\widetilde{ K}$, type $k$ works with probability $\gamma_k$ at threshold $b'_k+e$, and if $k\notin\widetilde{ K}$, type $k$ works with probability 1 at threshold $b'_k+e$. Let $D^1$ be the career value generated by the worker's strategy in this auxiliary equilibrium (under $R_n$). From all the arguments above, and due to the left-continuity of $\widetilde{R}_n$ and the continuity of $D(\cdot)$, we have that $D^0-D^1$ and $e$ both approximate zero as $\epsilon_n$ goes to zero. As a result, the fact that the auxiliary equilibrium is broken under $R_n$ implies $b'_k+e+D^1\neq\lambda_k$ for all $k$, which further implies $b'_k+D^0\neq\lambda_k$ for all $k$ whenever $\epsilon_n$ is small enough. Consequently, the candidate equilibrium cannot exist under $\widetilde{R}_n$. Finally, the last case we need to consider is one where some type $k$'s threshold bonus equals exactly some $\underline{b}_{k'}$. Likewise, by choosing a small $\epsilon_n$, we make sure that for each $k$, either $b'_k=\underline{b}_{k'}$ with some type $k'$, or $\widetilde{R}_n(b'_k)=R_n(b'_k)$. We again collect all the former category of types in $\widetilde{ K}$. For each $k\in\widetilde{ K}$, let $\gamma_k:=\sigma_k(b'_k)\frac{Q_k-R_n(\underline{b}_{k'}^+)}{R_n(\underline{b}_{k'})-R_n(\underline{b}_{k'}^+)}$. We again construct a similar auxiliary equilibrium under $R_n$ where each type $k$'s threshold bonus is $b'_k$, and if $k\in\widetilde{ K}$, type $k$ works with probability $\gamma_k$ at threshold $b'_k$, and if $k\notin\widetilde{ K}$, type $k$ works with probability 1 at threshold $b'_k$. Note that the career value generated by the worker's strategy in this auxiliary equilibrium (under $R_n$) is exactly $D^0$. Thus, the fact that the auxiliary equilibrium is broken under $R_n$ implies $b'_k+D^0\neq\lambda_k$ for all $k$. This condition further breaks the candidate equilibrium under $\widetilde{R}_n$. 

To conclude our proof of Lemma \ref{lemma-pure implementation}, we notice that the expected cost under $\widetilde{R}_n$ differs from that under $R_n$ only at those bonuses in the linearization intervals. However, once $\epsilon_n$ approximates zero as $n$ goes to infinity, this difference vanishes and the cost guarantee of $\widetilde{R}_n$ converges to $W^{\dagger}_Q$ as well.

One caveat is that Lemma \ref{lemma-pure implementation} does not guarantee that the robustly optimal incentive policy has $R^{\dagger}_Q(\underline{b}_k)=Q_k$ for all $k$, because $R^{\dagger}_Q$ is approximated by a policy sequence in the sense of weak convergence, so at its discontinuous points, the sequence may not converge pointwise. Instead, we have $R^{\dagger}_Q(\underline{b}_k)\geq Q_k$ for all $k$, because every incentive policy is nonincreasing and left-continuous. \\

Hereafter, we return to the proof of Proposition \ref{proposition-auxiliary problem-general}. We first derive the form of the objective function $V_Q(\cdot)$ in (\ref{auxiliary problem-general}). To do so, we fix an incentive policy $R$ that fully implements $Q$. Lemma \ref{lemma-pure implementation} suggests that it is without loss to consider a policy $R$ that fully implements $Q$ in the unique pure-strategy equilibrium where each type works with probability one at his threshold bonus $\underline{b}_k$. On the equilibrium path, when those bonuses in the interval $[\underline{b}_k,\underline{b}_{k-1})$ are realized, the types no lower than $k$ will work with probability one. Hence, the total probability of each such bonus being paid is $\sum_{j=1}^{k-1}\mu^0_jp_{0j}+\sum_{j=k}^K\mu^0_j(p_{0j}+p_j)$, which we denote as $r_k$. Recall that $F(b^-)=1-R(b)$. Therefore, let $Q_{K+1}=R(0)=1$, and the expected cost is
\begin{equation*}
    \begin{aligned}
        \sum_{k=1}^{K+1}r_k&\int_{\underline{b}_k}^{\underline{b}_{k-1}}bd(1-R(b))=\sum_{k=1}^{K+1}r_k\left(\Big[-bR(b)\Big]_{\underline{b}_k}^{\underline{b}_{k-1}}+\int_{\underline{b}_k}^{\underline{b}_{k-1}}R(b)db\right) \\
        &=\sum_{k=1}^{K+1}r_k\int_{\underline{b}_k}^{\underline{b}_{k-1}}R(b)db+\sum_{k=1}^{K+1}r_k[\underline{b}_{k}R(\underline{b}_{k})-\underline{b}_{k-1}R(\underline{b}_{k-1})] \\
        &=\underbrace{\sum_{k=1}^{K+1}r_k\int_{\underline{b}_k}^{\underline{b}_{k-1}}R(b)db}_{V_Q(R)}+\underbrace{\sum_{k=1}^{K+1}r_k(\underline{b}_{k}Q_{k}-\underline{b}_{k-1}Q_{k-1})}_{C_Q}.
    \end{aligned}
\end{equation*}
The first equality above results from integration by parts. The last equality above applies $R(\underline{b}_k)=Q_k$ for all $k$, because each type $k$ works with probability one at his threshold bonus $\underline{b}_k$. Hence, the expected cost equals $V_Q(R)+C_Q$, where $C_Q$'s value does not depend on $R$.
Finally, since the objective function is linear and thus continuous in $R$, it correctly measures the cost also at limits in the sense of Definition \ref{definition-solution concept-Q}. In particular, $W_Q^{\dagger}=V_Q(R^{\dagger}_Q)+C_Q$.

Now, to complete the proof of Proposition \ref{proposition-auxiliary problem-general}, it suffices to show the following claim: Every robustly optimal $R^{\dagger}_Q$ must be feasible in problem (\ref{auxiliary problem-general}). This is sufficient because, along with $W_Q^{\dagger}=V_Q(R^{\dagger}_Q)+C_Q$, this claim implies that the optimal value of (\ref{auxiliary problem-general}) offers a lower bound for $W^{\dagger}_Q$. Moreover, the claim makes sure that, for any solution to problem (\ref{auxiliary problem-general}), say $\overline{R}$, that can be approximated by a sequence $(R_n)_{n=1}^{\infty}$ in the sense of Definition \ref{definition-solution concept-Q}, this solution $\overline{R}$ must be robustly optimal. This is because, even though we have $W^{\dagger}_Q\geq V_Q(\overline{R})+C_Q$, the weak inequality must be equality since, otherwise, the employer could achieve a lower cost guarantee with $(R_n)_{n=1}^{\infty}$. Thus, $W^{\dagger}_Q=V_Q(\overline{R})+C_Q$ and $\overline{R}$ is robustly optimal.

To show the claim above, we take any robustly optimal $R^{\dagger}_Q$ and the sequence $(R_n)_{n=1}^{\infty}$ that approximates $R^{\dagger}_Q$ in the sense of Definition \ref{definition-solution concept-Q}. Lemma \ref{lemma-pure implementation} implies that it is without loss to assume $R_n(\underline{b}_k)=Q_k$ for all $n,k$. We know $R^{\dagger}_Q$ satisfies [FE]. Also, we have shown above that $R^{\dagger}_Q(\underline{b}_k)\geq Q_k$ for all $k$. Moreover, suppose $R^{\dagger}_Q(\underline{b}_k^+)>Q_k$. Then, there are a small $\epsilon>0$ and a large $n'$ such that for all $n>n'$, $R_n(\underline{b}_k+\epsilon)>Q_k$, a contradiction to $R_n(\underline{b}_k)=Q_k$ and $R_n(\underline{b}_k)\geq R_n(\underline{b}_k+\epsilon)$. Thus, $R^{\dagger}_Q$ satisfies [EK].

We show $R^{\dagger}_Q$ satisfies [EB-r] by contradiction. Suppose there is a $b'>\underline{b}_K$ such that [EB-r] is violated: $b'+D((R^{\dagger}_Q(b'+\lambda_k-\lambda_K))_{k=1}^K)<\lambda_K$. Then, by the continuity of $D(\cdot)$ and the left-continuity of $R^{\dagger}_Q$, there exists some $b^2$ slightly lower than $b'$ (or even $b'$ itself) such that $b^2+D((R^{\dagger}_Q(b^2+\lambda_k-\lambda_K))_{k=1}^K)<\lambda_K$ and all the threshold bonuses $b^2+\lambda_k-\lambda_K$ are continuous points of $R^{\dagger}_Q$. This implies that $R_n$ converges to $R^{\dagger}_Q$ pointwise at these threshold bonuses. Thus, the continuity of $D(\cdot)$ further admits a large $n$ such that $b^2+D((R_n(b^2+\lambda_k-\lambda_K))_{k=1}^K)<\lambda_K$. By the left-continuity of $R^{\dagger}_Q$ and $R_n$, we can always choose $b^2$ such that the threshold bonuses $b^2+\lambda_k-\lambda_K$ also are continuous points of $R_n$. On the other hand, for a large bonus $b^1>\max_{k,Q}\lambda_k-D(Q)$, we always have EB-r: $b^1+D((R_n(b^1+\lambda_k-\lambda_K))_{k=1}^K)>\lambda_K$. We also choose $b^1$ such that the threshold bonuses $b^1+\lambda_k-\lambda_K$ are continuous points of $R_n$. Next, we define the following correspondence, for all $b\in[b^2,b^1]$
\begin{equation*}
    \phi(b)=\{D(Q): Q_k=\gamma_kR_n(b+\lambda_k-\lambda_K)+(1-\gamma_k)R_n(b+\lambda_k-\lambda_K^+); \gamma_k\in[0,1]\},
\end{equation*}
which, fixing $b$ while varying all the $\gamma_k$'s, collects all the possible career values generated by the worker strategy that each type $k$ works with probability $\gamma_k$ at his threshold bonus $b+\lambda_k-\lambda_K$. One can check that the continuity of $D(\cdot)$ implies continuity for $\phi(\cdot)$. By construction, we have $\lambda_K-b^2>\phi(b^2)$ and $\lambda_K-b^1<\phi(b^1)$. Therefore, there must be some $b'\in(b^2,b^1)$ such that $\lambda_K-b'\in\phi(b')$, which means $R_n$ induces an equilibrium where the highest type's threshold bonus is $b'>b^2>\underline{b}_K$. This contradicts the assumption that $R_n$ fully implements $Q$. As a result, $R^{\dagger}_Q$ satisfies [EB-r] for all $b>\underline{b}_K$.

A similar argument shows that $R^{\dagger}_Q$ also satisfies [EB-l]. Notice that, for a small bonus $b^1<\min_{k,Q}\lambda_k-D(Q)$ (we assumed such $b^1>0$ exists in Section \ref{model}), $b^1$-[EB-l] always holds. Hence, by supposing [EB-l] is violated at some point and repeating the previous construction of an intermediate equilibrium, we obtain a contradiction. In summary, we have shown $R^{\dagger}_Q$ is feasible in (\ref{auxiliary problem-general}), further giving Proposition \ref{proposition-auxiliary problem-general}.

In fact, our arguments above also show that the closure (in terms of weak convergence) of $\mathcal{F}^{FI}(Q)$ is feasible in problem (\ref{auxiliary problem-general}). To see this, pick any $R^{\dagger}\in\cl(\mathcal{F}^{FI}(Q))$ and the sequence $(R_n)_{n=1}^{\infty}\subset\mathcal{F}^{FI}(Q)$ that weakly converges to $R^{\dagger}$. We also use (\ref{purification}) to adjust each $R_n$ while making it remain in $\mathcal{F}^{FI}(Q)$ and weakly converge to $R^{\dagger}$. Then, everything above goes through. \\

The last step is to prove Proposition \ref{proposition-auxiliary problem-private type}, which considers binary types $ K=\{H,L\}$ and full working $Q=(1,1)$. We start by showing problem (\ref{auxiliary problem-general}) is equivalent to problem (\ref{auxiliary problem-private type}) in this special case. First, the [EK] constraint for each $k$ degenerates to $R(\underline{b}_k)=1$ because $Q_k=1$, and [FE] guarantees $R(\underline{b}_k^+)\leq R(\underline{b}_k)=1$. Second, the objective function $V_Q(R)$ is thus equal to $\alpha+\beta\int_{\underaccent{\sim}{b}}^{\infty}R(s)ds$ with some positive $\alpha$ and $\beta$ since $R(b)=1$ for all $b\leq\underaccent{\sim}{b}=\underline{b}_L$. Third, [EB-r] is also satisfied at $\underline{b}_H$. In fact, we have $\underline{b}_H+D(R(\underline{b}_H+\lambda_L-\lambda_H),R(\underline{b}_H))=\underline{b}_H+D(1,1)=\lambda_H$. Finally, we show [EB-l] is implied by [EK] and [FE], so we can omit it from the problem. In particular, [EK] and [FE] require $R(b)=1$, for all $b<\underline{b}_H$, which means $b+D(R(\underline{b}_H+\lambda_L-\lambda_H),R(\underline{b}_H))=b+D(1,1)<\lambda_H$.

Recall that Proposition \ref{proposition-auxiliary problem-private type} also claims problem (\ref{auxiliary problem-private type}) produces the same set of solutions as the original problem (\ref{minimal cost guarantee-Q}) does. In the proof of Proposition \ref{main result-private type} below, we show problem (\ref{auxiliary problem-private type}) yields a unique solution that is also robustly optimal. That proof does not depend on the argument in this paragraph. Thus, the robustly optimal incentive policy must also be unique because, otherwise, there would be a different such incentive policy, which, by Proposition \ref{proposition-auxiliary problem-general}, is feasible in problem (\ref{auxiliary problem-private type}) while producing the same minimal cost guarantee. Nonetheless, this means problem (\ref{auxiliary problem-private type}) would not have a unique solution, forming a contradiction. Hence, the two problems induce the same set of solutions.

\subsection{Proof of Lemma \ref{lemma-greedy solution}}

Notice that $R^G$ must have a bounded support because, for those bonuses greater than $\max_{k,Q}\lambda_k-D(Q)$, the associated (\ref{equilibrium breaking-private type}) constraints cannot be binding, so $\supp(R^G)$ cannot contain these bonuses. Let the finite upper bound be $b^1=\max\supp(R^G)$. Thus, $R^G(b')=0$ for all $b'>b^1$. Since $b$-EB' holds if $R^G(b)<1$, for all $b'>b^1$, we have $b'+D(R^G(b'+\lambda_0),R^G(b'))=b'+D(0,0)\geq\lambda_H$, implying $b'\geq\lambda_H-D(0,0)=\accentset{\sim}{b}$. This can be possible only if $b^1\geq\accentset{\sim}{b}$. On the other hand, $b^1$-EB' binds because $b^1\in\supp(R^G)$, which means $\lambda_H=b^1+D(R^G(b^1+\lambda_0),R^G(b^1))=b^1+D(0,R^G(b^1))\geq b^1+D(0,0)\geq\accentset{\sim}{b}+D(0,0)=\lambda_H$ where the first inequality applies the assumption that $D$ is strictly increasing in $q_H$, and the second inequality comes from $b^1\geq\accentset{\sim}{b}$. Hence, the inequalities must be equalities, so $b^1=\accentset{\sim}{b}$ and $R^G(b^1)=0$.

Next, we show $R^G$ must be fully supported on an interval $[b_l,\accentset{\sim}{b}]$. To do this, suppose this is not true. That is, there exists some $b'\in\supp(R^G)$ such that $b'>b_l$ and, for a small $\epsilon>0$, $b'-\epsilon\notin\supp(R^G)$. Note that $b'>b_l$ implies $R^G(b')=R^G(b'-\epsilon)<1$ (otherwise, $R^G(b)=1$ for all $b\leq b'$, and $b_l\notin\supp(R^G)$). Hence, $b'$-EB' binds, while $(b'-\epsilon)$-EB' holds but does not bind. Formally
\begin{equation*}
    \begin{aligned}
        b'+D(&R^G(b'+\lambda_0),R^G(b'))=\lambda_H<b'-\epsilon+D(R^G(b'-\epsilon+\lambda_0),R^G(b'-\epsilon)) \\
        &<b'+D(R^G(b'+\lambda_0),R^G(b'-\epsilon))=b'+D(R^G(b'+\lambda_0),R^G(b')),
    \end{aligned}
\end{equation*}
which forms a contradiction. Here, the second inequality uses $R^G(b'+\lambda_0)\leq R^G(b'-\epsilon+\lambda_0)$ and the assumption that $D$ is strictly decreasing in $q_L$. Thus, $R^G$ is fully supported on an interval. Note that the same argument shows $R^G(b_l)=1$ because, otherwise, $R^G(b_l)=R^G(b_l-\epsilon)<1$ and $b_l\in\supp(R^G)$ while $b_l-\epsilon\notin\supp(R^G)$.

From above, we obtain $R^G(b)=0$ if $b\geq\accentset{\sim}{b}$, and $R^G(b)=1$ if $b\leq$ some $b_l$, and $R^G$ is strictly decreasing on $[b_l,\accentset{\sim}{b}]$. Furthermore, we show $R^G$ is also continuous on $[b_l,\accentset{\sim}{b}]$. Suppose this is not true, and we collect all the discontinuous points in $B^m$. For a small $\epsilon>0$, there exists $b'\in B^m$ such that $b'>(\sup B^m)-\epsilon$. In this case, $R^G$ is continuous at $b'+\lambda_0$. However, this cannot happen because $b'$-EB' and $(b'+0)$-EB' cannot both bind if the career value in $b'$-EB' jumps due to the discontinuity at $b'$ and the assumption that $D$ is strictly increasing in $q_H$. Consequently, $R^G$ is continuous. Finally, we claim $R^G$ is unique. Again, suppose this is not true, and for two different greedy incentive policies $\widetilde{R}^G$ and $\widehat{R}^G$, we collect all the points at which they differ in $B^d$. We have shown that $B^d\subset(b_l,\accentset{\sim}{b})$. For a small $\epsilon>0$, there exists $b'\in B^d$ such that $b'>(\sup B^d)-\epsilon$. In this case, $q':=\widetilde{R}^G(b'+\lambda_0)=\widehat{R}^G(b'+\lambda_0)$ and $b'$-EB' binds for both policies. However, this means $D(q,\widetilde{R}^G(b'))=D(q,\widehat{R}^G(b'))$ while $\widetilde{R}^G(b')\neq\widehat{R}^G(b')$. The assumption that $D$ is strictly increasing in $q_H$ forbids such a situation. In summary, $R^G$ is unique. The final thing to show is $b_l<\underaccent{\sim}{b}$. One can see this by using the binding $b_l$-EB', which gives $\lambda_H=b_l+D(R^G(b_l+\lambda_0),R^G(b_l))>b_l+D(1,1)=b_l+\lambda_L-\underaccent{\sim}{b}$. The inequality here applies $R(b_l)=1$ and the fact that $R^G$ is strictly decreasing on an interval above $b_l$, so $R(b_l+\lambda_0)<R(b_l)$. This simply indicates that $b_l<\underaccent{\sim}{b}-(\lambda_L-\lambda_H)<\underaccent{\sim}{b}$.

\subsection{Proof of Proposition \ref{main result-private type}}

By the definition of the greedy incentive policy $R^G$ and Lemma \ref{lemma-greedy solution}, we observe that Proposition \ref{main result-private type} states that the robustly optimal incentive policy $F^{\dagger}$ is associated with the tail incentive policy $R^{\dagger}$ given by
\begin{equation}\label{R dagger}
    R^{\dagger}(b)=\begin{cases}
        1 &\text{if }b\leq\underaccent{\sim}{b}; \\
        R^G(b) &\text{otherwise.}
    \end{cases}
\end{equation}
We first show that $R^{\dagger}$ is the unique solution to problem (\ref{auxiliary problem-general}), which, in this case, coincides with problem (\ref{auxiliary problem-private type}) where (\ref{equilibrium keeping-private type}) and (\ref{equilibrium breaking-private type}) correspond to [EK] and [EB-r], respectively. Since [EB-l] is not important in the binary-type case, we also use [EB] as a shorthand for [EB-r]. Then, we apply Proposition \ref{proposition-auxiliary problem-general} to verify that this solution is robustly optimal. In the proof of Proposition \ref{proposition-auxiliary problem-private type}, we already showed this also implies the uniqueness of robustly optimal incentive policy.

First, denote by $V(R):=\int_{\underaccent{\sim}{b}}^{\infty}R(s)ds$ the objective function in problem (\ref{auxiliary problem-private type}), which differs from the objective in problem (\ref{auxiliary problem-general}) only by a constant. Suppose there is another $\widetilde{R}$ feasible in problem (\ref{auxiliary problem-general}) such that $V(\widetilde{R})\leq V(R^{\dagger})$ and $\widetilde{R}\neq R^{\dagger}$. The [EK] constraint in problem (\ref{auxiliary problem-general}) then implies $\widetilde{R}(b)=R^{\dagger}(b)=1$ for all $b\leq\underaccent{\sim}{b}$, so their difference must lie in some bonuses greater than $\underaccent{\sim}{b}$. Note that this can happen only if $\accentset{\sim}{b}>\underaccent{\sim}{b}$ since, otherwise, $R^{\dagger}(b)=R^G(b)=0$ for all $b>\underaccent{\sim}{b}$ and one could not find a different policy that yields weakly lower expected cost. By assumption, there must be a bonus $b$ such that $\widetilde{R}(b)<R^{\dagger}(b)$, and we collect all such bonuses in $B^d$. We must have $B^d\subset(\underaccent{\sim}{b},\accentset{\sim}{b})$ since $R^{\dagger}(b)=R^G(b)=0$ at a higher $b$. Therefore, we take a small $\epsilon>0$ and find the $b'\in B^d$ such that $b'>(\sup B^d)-\epsilon$. This implies $\widetilde{R}(b')<R^{\dagger}(b')$ while $\widetilde{R}(b'+\lambda_0)\geq R^{\dagger}(b'+\lambda_0)$. However, $b'$-EB' binds at $R^{\dagger}$ by the definition of $R^G$, which gives $\lambda_H=b'+D(R^{\dagger}(b'+\lambda_0),R^{\dagger}(b'))>b'+D(\widetilde{R}(b'+\lambda_0),\widetilde{R}(b'))$. Hence, $\widetilde{R}$ violates $b'$-EB' and thus cannot be feasible in problem (\ref{auxiliary problem-private type}). As a result, $R^{\dagger}$ is the unique solution to problem (\ref{auxiliary problem-private type}).

To apply Proposition \ref{proposition-auxiliary problem-general}, we construct a sequence $(R_n)_{n=1}^{\infty}$ that approximates $R^{\dagger}$ in the sense of Definition \ref{definition-solution concept-Q} (in this case, Definition \ref{definition-solution concept}). We take a sequence of small numbers $(\epsilon_n)_{n=1}^{\infty}$ that converges to zero. Then, we construct each $R_n$ by setting $R_n(b)=R^{\dagger}(b-\epsilon_n)$, which amounts to shifting $R^{\dagger}$ to the right by $\epsilon_n$. Since $R_n$ converges to $R^{\dagger}$ pointwise, it also weakly converges to $R^{\dagger}$. It suffices to show $R_n$ fully implements full working. Since $R_n(\underline{b}_k)=1$ for both $k$, full working is induced in one equilibrium. For all $b\geq\underline{b}_H+\epsilon_n$, the fact that $R^{\dagger}$ satisfies [EB] gives $b-\epsilon_n+D(R^{\dagger}(b+\lambda_0-\epsilon_n),R^{\dagger}(b-\epsilon_n))\geq\lambda_H$. Hence, under $R_n$, we have $b+D(R_n(b+\lambda_0),R_n(b))=b+D(R^{\dagger}(b+\lambda_0-\epsilon_n),R^{\dagger}(b-\epsilon_n))\geq\lambda_H+\epsilon_n>\lambda_H$. For all $b\in(\underline{b}_H,\underline{b}_H+\epsilon_n)$, we have $b+D(R_n(b+\lambda_0),R_n(b))=b+D(1,1)>\underline{b}_H+D(1,1)=\lambda_H$. In summary, every undesirable candidate equilibrium where types $k=H,L$ work with probability one at their respective threshold bonuses cannot exist. Next, we consider the mixed-strategy candidate equilibria. We start with those where the high type randomizes at the threshold bonus $\underline{b}_L+\epsilon_n$, where $R_n$ has a mass point. In this case, the assumption that $D$ is strictly increasing in $q_H$ guarantees that the career value generated by any interior working probability is strictly higher than that in the candidate equilibrium with a high-type threshold bonus slightly higher than $\underline{b}_L+\epsilon_n$, so the fact that the latter equilibrium cannot exist also breaks the former. The remaining case includes those candidate equilibria where the low type randomizes at threshold $\underline{b}_L+\epsilon_n$. Likewise, the assumption that $D$ is strictly decreasing in $q_L$ guarantees that the career value generated by any interior working probability is strictly higher than that in the candidate equilibrium with a low-type threshold bonus slightly lower than $\underline{b}_L+\epsilon_n$. Thus, such an equilibrium cannot exist. Consequently, $R_n$ fully implements full working, and Proposition \ref{proposition-auxiliary problem-general} hence suggests that $R^{\dagger}$ is robustly optimal.

\subsection{Proof of Proposition \ref{main result-Q} and Proposition \ref{proposition-implementability}}

We begin by showing one side of Proposition \ref{proposition-implementability}: For $Q_H\geq Q_L$, if $Q_L\leq R^G(\underline{b}_L)$ and $\underline{b}_L\in\supp(R^G)$, then $Q$ is not implementable. By Lemma \ref{lemma-greedy solution}, $\underline{b}_L\in\supp(R^G)=[b_l,\accentset{\sim}{b}]$ implies $[\underline{b}_L,\accentset{\sim}{b}]$ is nonempty. Suppose $Q$ is fully implemented by some $R$. The proof of Lemma \ref{lemma-pure implementation} shows it is without loss to choose $R$ such that $R(\underline{b}_L)=Q_L\leq R^G(\underline{b}_L)$. Therefore, if we collect all bonuses $b\in[\underline{b}_L,\accentset{\sim}{b}]$ such that $R(b)\leq R^G(b)$ in $B^l$, $B^l$ is not empty. We take a small $\epsilon>0$ and find some $b'\in B^l$ such that $b'>(\sup B^l)-\epsilon$. This means $R(b'+\lambda_0)\geq R^G(b'+\lambda_0)$ because either $b'+\lambda_0\leq\accentset{\sim}{b}$ but $b'+\lambda_0\notin B^l$, or $b'+\lambda_0>\accentset{\sim}{b}$ so $R^G(b'+\lambda_0)=0$. By the assumption that $D$ is strictly increasing in $q_H$ and strictly decreasing in $q_L$, and by the construction of $R^G$, we have $b'+D(R(b'+\lambda_0),R(b'))\leq b'+D(R^G(b'+\lambda_0),R^G(b'))=\lambda_H$. In addition, the proof of Proposition \ref{proposition-auxiliary problem-general} implies that such $R\in\mathcal{F}^{FI}(Q)$ must be feasible in problem (\ref{auxiliary problem-general}), so $b'$-[EB] gives: $b'+D(R(b'+\lambda_0),R(b'))\geq\lambda_H$. Combining both inequalities above yields $b'+D(R(b'+\lambda_0),R(b'))=\lambda_H$, which means there is an equilibrium where the high type's threshold bonus is $b'\geq\underline{b}_L>\underline{b}_H$. This forms a contradiction and thus $Q$ is not implementable. \\

Next, we show Proposition \ref{main result-Q} by first demonstrating that the stated policy, $R^{\dagger}_Q$, given by (\ref{ROWP-Q}) is the unique solution to problem (\ref{auxiliary problem-general}) and then applying Proposition \ref{proposition-auxiliary problem-general} to verify that it is robustly optimal. To repeat the argument at the end of Proposition \ref{proposition-auxiliary problem-private type}'s proof, we want to show $R^{\dagger}_Q$ is the unique robustly optimal incentive policy. Our analysis above implies that we must have either $\underline{b}_L>\accentset{\sim}{b}$ or $Q_L>R^G(\underline{b}_L)$ for an implementable $Q$. In fact, $\underline{b}_L>\accentset{\sim}{b}$ implies $Q_L\geq R^G(\underline{b}_L)$, so we always have the latter.

To start with, we show $R^{\dagger}_Q$ is feasible in problem (\ref{auxiliary problem-general}). First, the construction of $R^G$ and the definition of $\underline{b}_k$ give $\underline{b}_H+D(R^G(\underline{b}_L),R^G(\underline{b}_H))=\lambda_H=\underline{b}_H+D(Q_L,Q_H)$. Hence, $Q_L\geq R^G(\underline{b}_L)$ and the assumption that $D$ is strictly increasing in $q_H$ and strictly decreasing in $q_L$ together imply $Q_H\geq R^G(\underline{b}_H)$. (\ref{ROWP-Q}) thus gives $R^{\dagger}_Q(\underline{b}_k)=Q_k$ for both $k$, which satisfies [EK]. Second, for all $b>\underline{b}_H$, $R^{\dagger}_Q$ satisfies [EB-r] because $b+\lambda_0>\underline{b}_L$, meaning $R^{\dagger}_Q(b+\lambda_0)=R^G(b+\lambda_0)$. The maximum expression of (\ref{ROWP-Q}) further gives $R^{\dagger}_Q(b)\geq R^G(b)$. Thus, $b+D(R^{\dagger}_Q(b+\lambda_0),R^{\dagger}_Q(b))\geq b+D(R^G(b+\lambda_0),R^G(b))=\lambda_H$, satisfying $b$-[EB-r]. Lastly, for all $b<\underline{b}_H$, $R^{\dagger}_Q$ satisfies [EB-l] since $b+\lambda_0<\underline{b}_L$, implying $R^{\dagger}_Q(b+\lambda_0)\geq R^{\dagger}_Q(\underline{b}_L)=Q_L$. Hence, $R^{\dagger}_Q(b)=Q_H$ and the assumption that $D$ is strictly decreasing in $q_L$ together yield $b+D(R^{\dagger}_Q(b+\lambda_0),R^{\dagger}_Q(b))\leq b+D(Q_L,Q_H)<\underline{b}_H+D(Q_L,Q_H)=\lambda_H$, satisfying $b$-[EB-l]. Of course, [FE] holds for $R^{\dagger}_Q$ (as $R^G_Q$ and $R^P_Q$ satisfy [FE]).

Knowing $R^{\dagger}_Q$ is feasible, we now show $R^{\dagger}_Q$ is the unique solution to problem (\ref{auxiliary problem-general}). Suppose there is a different feasible $\widetilde{R}$ that produces weakly lower expected cost than $R^{\dagger}_Q$ does. In problem (\ref{auxiliary problem-general}), the fact that the objective function is linear and the coefficients $r_k$ are strictly positive implies $B^s:=\{b\geq0:\widetilde{R}(b)<R^{\dagger}_Q(b)\}$ is nonempty. In addition, [EK] requires $\widetilde{R}\geq R^K_Q$, and hence the fact that $R^{\dagger}_Q(\underline{b}_k)=Q_k$ for both $k$ means whenever $b\in B^s$, we have $R^{\dagger}_Q(b)=R^G(b)$. We take a small $\epsilon>0$ and find $b'\in B^s$ such that $b'>(\sup B^s)-\epsilon$, so that $\widetilde{R}(b'+\lambda_0)\geq R^{\dagger}_Q(b'+\lambda_0)\geq R^G(b'+\lambda_0)$. Then, $\widetilde{R}(b')<R^{\dagger}_Q(b')=R^G(b')$ and the assumption that $D$ is strictly increasing in $q_H$ and strictly decreasing in $q_L$ together imply $b'+D(\widetilde{R}(b'+\lambda_0),\widetilde{R}(b'))<b'+D(R^G(b'+\lambda_0),R^G(b'))=\lambda_H$, a contradiction to $b'$-[EB-r]. In summary, $R^{\dagger}_Q$ is the unique solution to problem (\ref{auxiliary problem-general}).

To apply Proposition \ref{proposition-auxiliary problem-general}, we need to construct a sequence $(R_n)_{n=1}^{\infty}$ that approximates $R^{\dagger}_Q$ in the sense of Definition \ref{definition-solution concept-Q}. First, we deal with the following two cases: either $Q_H=Q_L$ or $Q_L=0$. From (\ref{ROWP-Q}), $R^{\dagger}_Q$ in both cases contains at most one mass point. In fact, the $Q_L=0$ case does not produce any mass point as $\underline{b}_H+D(0,R^{\dagger}_Q(\underline{b}_H))=\underline{b}_H+D(0,Q_H)=\lambda_H=\underline{b}_H+D(0,R^G(\underline{b}_H))$. Whereas, the $Q_H=Q_L>0$ case contains one mass point at $\underline{b}_L$ because, at $b'$ such that $R^G(b')=Q_H=Q_L(>R^G(b'+\lambda_0))$, we have $\lambda_H=b'+D(R^G(b'+\lambda_0),R^G(b'))>b'+D(Q_L,Q_H)=b'+\lambda_L-\underline{b}_L$, which gives $b'<\underline{b}_L-(\lambda_L-\lambda_H)$ and hence $Q_L>R^G(\underline{b}_L)$. Thus, the associated part of Proposition \ref{main result-private type}'s proof where we essentially constructed $R_n(b)=R^{\dagger}_Q(b-\epsilon_n)$ for small $\epsilon_n>0$ converging to 0 still works here. Note that, in the $Q_L=0$ case, since $Q_L\leq R^G(\underline{b}_L)$, an implementable $Q$ must have $\underline{b}_L>\accentset{\sim}{b}$, which ensures that such construction maintains $R_n(\underline{b}_k)=Q_k$ for both $k$ (namely [EK]) and thus also works in this case. Second, we consider the last case $Q_H>Q_L>0$. Note that we must have $Q_L>R^G(\underline{b}_L)$ since, otherwise, $R^G(\underline{b}_L)\geq Q_L>0$ implies $\underline{b}_L\leq\accentset{\sim}{b}$, which, as we have shown, makes $Q$ not implementable. We also have shown $D(R^G(\underline{b}_L),R^G(\underline{b}_H))=D(Q_L,Q_H)$, which further implies $Q_H>R^G(\underline{b}_H)$. In summary, $R^{\dagger}_Q$ contains two mass points at $\underline{b}_k$ for both $k$. We therefore take a sequence of small positive numbers $(\epsilon_n)_{n=1}^{\infty}$ that converges to 0, so as to construct each $R_n$ in the following way
\begin{equation*}
    R_n=\begin{cases}
        R^{\dagger}_Q(b-\epsilon_n) & \text{ if }b\geq\underline{b}_L+\epsilon_n; \\
        R^{\dagger}_Q(b-2\epsilon_n) & \text{ if }b<\underline{b}_L+\epsilon_n.
    \end{cases}
\end{equation*}
Due to the presence of two mass points, by choosing a small $\epsilon_n$, we maintain $R_n(\underline{b}_k)=Q_k$ for both $k$, and $R_n$ satisfies [FE] and induces $Q$ in one equilibrium. We then show $R_n$ fully implements $Q$. First, for all $b\geq\underline{b}_L+\epsilon_n$, we have $b+D(R_n(b+\lambda_0),R_n(b))=\epsilon_n+(b-\epsilon_n)+D(R^{\dagger}_Q(b+\lambda_0-\epsilon_n),R^{\dagger}_Q(b-\epsilon_n))\geq\epsilon_n+\lambda_H>\lambda_H$. For all $b\in[\underline{b}_H+\epsilon_n,\underline{b}_L+\epsilon_n)$, we have $b+D(R_n(b+\lambda_0),R_n(b))=b+D(R^{\dagger}_Q(b+\lambda_0-\epsilon_n),R^{\dagger}_Q(b-2\epsilon_n))\geq b+D(R^{\dagger}_Q(b+\lambda_0-\epsilon_n),R^{\dagger}_Q(b-\epsilon_n))=\epsilon_n+(b-\epsilon_n)+D(R^{\dagger}_Q(b+\lambda_0-\epsilon_n),R^{\dagger}_Q(b-\epsilon_n))\geq\epsilon_n+\lambda_H>\lambda_H$. For all $b<\underline{b}_H+\epsilon_n$, we have $b+D(R_n(b+\lambda_0),R_n(b))=2\epsilon_n+(b-2\epsilon_n)+D(R^{\dagger}_Q(b+\lambda_0-2\epsilon_n),R^{\dagger}_Q(b-2\epsilon_n))\geq\epsilon_n+\lambda_H>\lambda_H$. Therefore, as in the proof of Proposition \ref{main result-private type}, now we just need to break all the mixed equilibria. Two possibilities arise: The high type randomizes at threshold $\underline{b}_H+2\epsilon_n$ (the low type's threshold $\underline{b}_L+2\epsilon$ is a continuous point of $R_n$); or, the low type randomizes at his threshold $\underline{b}_L+\epsilon_n$ (the high type's threshold $\underline{b}_H+\epsilon_n$ is a continuous point of $R_n$). As in the proof of Proposition \ref{main result-private type}, the career value generated by any threshold behavior in the former equilibrium is strictly higher than that in the candidate equilibrium with the high-type threshold slightly higher than $\underline{b}_H+2\epsilon_n$, while the latter case's career value is strictly higher than in the candidate equilibrium with a low-type threshold slightly lower than $\underline{b}_L+\epsilon_n$. Since the two candidate equilibria are broken by $R_n$, so are the mixed ones. In summary, Proposition \ref{proposition-auxiliary problem-general} suggests that $R^{\dagger}_Q$ is robustly optimal.

Notice that we have already shown that whenever $Q_L>R^G(\underline{b}_L)$ or $\underline{b}_L>\accentset{\sim}{b}$, $Q$ can be implemented by some $R_n$ constructed above. As a result, we prove Proposition \ref{proposition-implementability} at the same time.

\subsection{Proof of Proposition \ref{proposition-criterion for contract dispersion-general}}

As in our discussion in Section \ref{strategic uncertainty}, full working forms an equilibrium if the lowest type is willing to work conditional on the career value $D(1,...,1)$. Thus, the partial-implementation bonus here is still $\underaccent{\sim}{b}$. Next, we consider the deterministic robust bonus $b^R$. To keep the desired equilibrium, we must have $b^R\geq\underaccent{\sim}{b}$. To break all the undesirable equilibria, notice that given a deterministic contract with bonus $b'$, such a candidate equilibrium must have some $k\in K$ and some $q\in[0,1]$ such that the total working probability profile is $Q_{k,q}=(0,...,0,q,1,...,1)$ with $k-1$ zeros. If $q\in(0,1)$, the type $k$ worker must be indifferent between working and shirking. To break an equilibrium with $k$ and $q\in(0,1)$, we must have $b'\neq\lambda_k-D(Q_{k,q})$; whereas, to break an equilibrium with $k$ and $q=0$ (or $k+1$ and $q=1$), we need $b'>\lambda_k-D(Q_{k,0})$ or $b'<\lambda_{k+1}-D(Q_{k+1,1})$. A subset of these conditions is $b'\neq\lambda_k-D(Q_{k,q})$ for all $k$ and $q\in[0,1]$. To check when this subset will be satisfied, we define $B^m=\{\lambda_k-D(Q_{k,q}):k\in K,q\in[0,1]\}$. Due to the continuity of $D(\cdot)$, for each $k$, the set $B^m_k:=\{\lambda_k-D(Q_{k,q}):q\in[0,1]\}$ is an interval. Also, for all $k<K$, $Q_{k,0}=Q_{k+1,1}$, and thus $\lambda_k-D(Q_{k,0})\in B^m_k\cap B^m_{k+1}$. As a result, $B^m$ is also an interval. The fact that $b'\geq\underaccent{\sim}{b}$ and $\underaccent{\sim}{b}=\lambda_1-D(Q_{1,1})\in B^m$ implies $b'>\max B^m$. Notice that the conditions for breaking the two types of undesirable equilibria mentioned above are already implied by $b'>\max B^m\geq\lambda_k-D(Q_{k,q})$ for all $k$ and $q\in[0,1]$, so the deterministic robust bonus is simply the infimum bonus satisfying this, yielding $b^R=\max B^m$. Finally, we show $\max B^m=\accentset{\sim}{b}$. This is true because, fixing each $k$, the fact that $D(\cdot)$ is quasi-concave in $q_k$ implies that $\lambda_k-D(Q_{k,q})$ is maximized (namely $D(Q_{k,q})$ is minimized) at either $q=0$ or $1$. However, if $k>1$, $\lambda_{k-1}-D(Q_{k-1,0})>\lambda_k-D(Q_{k,1})$, so $q=1$ cannot form the maximizer in $B^m$. In summary, only two such cases can be the maximizer in $B^m$: some $k$ and $q=0$, or $q=1$ with $k=1$. The former corresponds to $\max_{k\in K}\lambda_k-D(0,...,0,1,...,1)$ with $k$ zeros (in the definition of $\accentset{\sim}{b}$) while the latter corresponds to $\underaccent{\sim}{b}$. The analysis above further indicates that, if $\accentset{\sim}{b}=\underaccent{\sim}{b}$, the partial-implementation bonus coincides with the deterministic robust bonus, so it suffices to fully implement full working and thus it is a (unique) robustly optimal incentive policy. As a result, dispersion is not robustly optimal.

In the opposite direction, we now consider the case $\accentset{\sim}{b}>\underaccent{\sim}{b}$. We construct a new incentive policy with dispersion that improves upon the deterministic robust bonus $\accentset{\sim}{b}$, which shows no dispersion cannot be robustly optimal and therefore concludes the proof. To do so, we take three positive numbers $e$, $\epsilon<1$, and $\delta$. The new incentive policy, denoted as $F'$, assigns probability $\epsilon$ to $\accentset{\sim}{b}+e$ and the rest of probability to $\accentset{\sim}{b}+e-\delta$. The expectation of $F'$ is $\accentset{\sim}{b}+e-\delta(1-\epsilon)$, so by choosing $e<\delta(1-\epsilon)$, we obtain an improvement. The remaining job is to construct a certain $\epsilon$ (that depends on $\delta$ but not $e$) to ensure that $F'$ fully implements full working. By choosing $\delta<\min_{k<K}\lambda_k-\lambda_{k+1}$, we make sure that, in each candidate equilibrium, there is at most one type whose threshold lies between the two supporting bonus levels, which means the total working probability profile must be $Q_{k,q}$ for some $k$ and $q\in[0,1]$. Let $\widetilde{ K}=\argmax_{k\in K}\lambda_k-D(Q_{k,0})$ collect all the types $k$ that attain $\accentset{\sim}{b}$ in (\ref{criterion for contract dispersion-general}). Let $E^d=\{(k,q)\in K\times[0,1]:\lambda_k-D(Q_{k,q})\geq\accentset{\sim}{b}-\delta\}$ collect all the candidate equilibria where the lower bonus $\accentset{\sim}{b}+e-\delta$ is not high enough to induce more work from type $k$. All the other candidate equilibria (without full working) are broken because even the low bonus $\accentset{\sim}{b}+e-\delta$ is sufficiently rewarding. In fact, we have $E^d\subset\cup_{k^*\in\widetilde{ K}}\{(k^*,q):q\in[0,\epsilon_{k^*}]\}$ with some $(\epsilon_{k^*})_{k^*\in\widetilde{ K}}$. This is because $\lambda_{k+1}-D(Q_{k+1,1})<\lambda_k-D(Q_{k,0})$ for all $k<K$, so we can always choose a small $\delta$ to ensure that these hard-to-break equilibria in $E^d$ are only those close to and on the right-hand side of the hardest-to-break equilibria $(k^*,0)$. In addition, for $k^*>1$, $\lambda_{k^*-1}-D(Q_{k^*-1,0})>\lambda_{k^*}-D(Q_{k^*,1})$, while if $k^*=1$, we have $\lambda_{k^*}-D(Q_{k^*,1})=\underaccent{\sim}{b}<\accentset{\sim}{b}$. Both cases suggest that $D(Q_{k^*,1})>D(Q_{k^*,0})$. Hence, the continuity of $D(\cdot)$ means, for a small $\delta$, we can always have $\epsilon_{k^*}<1$. We thus choose $\epsilon$ such that $1>\epsilon>\epsilon_{k^*}$ for all $k^*\in\widetilde{ K}$. This construction makes sure that each candidate equilibrium $(k,q)$ in $E^d$ must be such that the type $k$ worker's threshold bonus is no less than the higher bonus $\accentset{\sim}{b}+e$ (otherwise, $q>\epsilon$), which is even higher than the deterministic robust bonus. As a result, all these candidate equilibria are broken, and hence $F'$ fully implements full working.

\end{document}